%% file: edpf_ver2.tex
\documentclass[10pt]{extarticle}
\usepackage{amsmath,amsthm,amssymb,amsbsy,amsopn,mathtools,comment}
\usepackage{graphicx,color}
\usepackage[bookmarks=true,bookmarksnumbered=true,bookmarkstype=toc]{hyperref}
\usepackage{mathptmx,bbm}
\usepackage{mathtools}
\usepackage{lineno}
\usepackage[dvipsnames]{xcolor}
\usepackage{xcolor, soul}

\def\argmin{\mathop{\rm argmin}}

\theoremstyle{definition}

\makeatletter
\@addtoreset{equation}{section}
\makeatother

\title{The Gerber-Shiu discounted penalty function: \textcolor{black}{A review} from practical perspectives\footnotetext[0]{
This work was partially supported by JSPS Grants-in-Aid for Scientific Research 19H01791, 20K03758, 20K22301, 21K03347 and 21K03358.}}
\author{\sc Yue He\footnote{Email address: y.he@maths.usyd.edu.au. Postal Address: School of Mathematics and Statistics, The University of Sydney, Australia.}
,\, Reiichiro Kawai\footnote{Corresponding author. 
Email address: raykawai@g.ecc.u-tokyo.ac.jp. Postal address: Graduate School of Arts and Sciences / Mathematics and Informatics Center, The University of Tokyo, Japan, and School of Mathematics and Statistics, The University of Sydney, Australia.},\, Yasutaka Shimizu\footnote{Email address: shimizu@waseda.jp. Postal address: Department of Applied Mathematics, Waseda University, Japan.}\,\, and Kazutoshi Yamazaki\footnote{Email address: k.yamazaki@uq.edu.au. Postal address: School of Mathematics and Physics, The University of Queensland, Australia.}}
\date{}

\begin{document}

\maketitle

\input{sections_ver2x/edpf_abst_section1.tex}
\input{sections_ver2x/edpf_section2.tex}
\input{sections_ver2x/edpf_section3.tex}
\input{sections_ver2x/edpf_section4.tex}

\input{sections_ver2x/edpf_section5.tex}
\input{sections_ver2x/edpf_concludingremarks.tex}

\section*{Declaration of competing interest}

None declared.

\small
\bibliographystyle{abbrv}
\bibliography{edpfbibliography}

\end{document}

%% file: sections_ver2x/edpf_abst_section1.tex

\begin{abstract}
\noindent 
The Gerber-Shiu function provides a unified framework for the evaluation of a variety of risk quantities.
Ever since its establishment, it has attracted constantly increasing interests in actuarial science, whereas the conventional research has been focused on finding analytical or semi-analytical solutions, either of which is rarely available, except for limited classes of penalty functions on rather simple risk models.
In contrast to its great generality, the Gerber-Shiu function does not seem sufficiently prevalent in practice, largely due to a variety of difficulties in numerical approximation and statistical inference.
To enhance research activities on such implementation aspects, we provide a \textcolor{black}{comprehensive} review of existing formulations and underlying surplus processes, as well as an extensive survey of analytical, semi-analytical and asymptotic methods for the Gerber-Shiu function, which altogether shed fresh light on its numerical methods and statistical inference for further developments.
On the basis of an ambitious collection of 235 references, the present survey can serve as an insightful guidebook to model and method selection from practical perspectives as well.

\vspace{0.3em}
\noindent {\it Keywords:} Gerber-Shiu function; Laplace transform; integro-differential equations; series expansions; scale function.

\vspace{0.3em}
\noindent {\it JEL classification:} C14, C63, G22.

\noindent {\it 2020 Mathematics Subject Classification:} 91G05, 60G40, 91G60, 91G70.
\end{abstract}

\tableofcontents

\section{Introduction}

The Gerber-Shiu function, also referred to as the expected discounted penalty function, was first constructed back in 1998 by Gerber and Shiu \cite{Gerber_Shiu_1998}. 
It is a functional of the time of ruin, that is, the first time when the surplus of the insurance company reach below a threshold level, as well as the surplus prior to ruin, and the deficit at ruin.
Ever since its establishment, the Gerber-Shiu function has attracted constantly increasing interests in actuarial science due to its generality in representing numerous ruin-related quantities, including the ultimate ruin probability, the Laplace transform of the time of ruin and the joint tail distribution of surplus and deficit (Section \ref{section formulation}). 
In addition, a wide variety of its variants have been developed (Section \ref{section Extended Gerber-Shiu function}), for instance, via generalizations on the penalty function (Section \ref{sec_generalization_w}) and the discount factor (Section \ref{Section Generalization of the discount factor}), as well as with a finite-time horizon (Section \ref{section Finite time horizon}). 

In the literature, a variety of stochastic processes have been employed in the context of the Gerber-Shiu function in modeling the temporal evolution of the surplus of an insurance company (Section \ref{section Surplus processes}).
Among those, the first model is built on compound Poisson processes, also known as the Cram\'er-Lundberg model or classical risk model 
(Section \ref{section Cl model}). 
The Cram\'er-Lundberg model has then been generalized in many different directions ever since, so as to better capture stylized features of the surplus process in the real world, such as L\'evy risk models
(Section \ref{section Levy risk model}) often with diffusion perturbations (Section \ref{section_diffusion_perturbation}), the renewal risk model, also known as the Sparre-Andersen model (Section \ref{sec_Sparre-Andersen}), whose number of claims is modeled by a renewal process with independent and identically distributed interclaim times,
and Markov additive processes (Section \ref{subsubsection markov additive processes}) that offer better tractability of renewal risk models with phase-type interarrival times, 
along with suitable additional components (Section \ref{subsubsection surplus processes with additional components}).
Moreover, with a view towards real-world applications, more advanced structural features have also been proposed in specialized manners (Section \ref{subsection formulations specialized for applications}), for instance, in terms of the drift component, such as interest earnings, debit or tax payments, and level dependence structure due to dividend payments (Section \ref{Drift component}), and for modeling credit default swaps (Section \ref{section CDS}), capital injection (Section \ref{subsubsection capital injection}) and risk measures (Section \ref{subsubsection risk measures}).

As such, there exist quite a wide variety of developments in the literature on the Gerber-Shiu function, yet with a rather restricted focus on finding its analytical or semi-analytical expressions. 
Under somewhat strict assumptions on each component of the problem formulation, its analytical or semi-analytical expression may be available, for instance, by inverting its Laplace transform (Section \ref{section closed - Laplace transform}), or by solving defective renewal equations (Section \ref{section closed - Defective renewal equation}) and the integro-differential equation with suitable boundary conditions (Section \ref{section closed - Solve PIDE}). 
Alternatively, the Gerber-Shiu function can be expressed in terms of the scale function of certain risk surplus processes (Section \ref{section closed - Scale functions}), recursive representations (Section \ref{section closed -  Recursive}), and integral representations with respect to suitable probability measures (Section \ref{section closed -  Derivation of the Gerber-Shiu measure}). 
Furthermore, its asymptotic expressions have been derived for a wide range of claim size distributions when analytic expressions are not easily available (Section \ref{subsubsection Asymptotic results}). 

It is however a long-standing consensus that analytical representations are rarely available for the Gerber-Shiu function at convenience, unless risk models and claim size distributions are selected from limited classes. 
Even with semi-analytical expressions, for instance, in terms of integrals or infinite series, suitable numerical approximation is required for the evaluation at an implementation stage.
Nevertheless, numerical methods on the Gerber-Shiu function do not seem to have been adequately explored in the literature.
To the best of our knowledge, there exist only a handful of developments in the context of the Gerber-Shiu function: the numerical inversion of the Laplace transform \cite{library979190, kuznetsov_morales_2014} (Section \ref{section Laplace Transform}), finite truncation of the Fourier expansion \cite{CHAU2015170, Li_Shi_Yam_Yang_2021, xie_zhang_2021, xie_zhang_2022} (Section \ref{section Fourier-cosine expansion}), the numerical solution to the integro-differential equation with suitable boundary conditions \cite{DIKO2011126, dong_zhao_2012, insp, risks8010024, zhou_xiao_deng_2015, zhuo_yang_chen_2018} (Section \ref{section Approximate solution of integro-differential equation}), and simulation via random number generation \cite{COSSETTE_2015, ferreiro-castilla_van_schaik_2015} (Section \ref{section simulation}).
As is always the case, the method needs to be carefully chosen for each problem setting, because each of those methods above has both advantages and disadvantages with different technical prerequisites for implementation and convergence (Section \ref{subsection discussion}).
In contrast to other physical and mathematical sciences with an abundance of numerical methods, this very short list acts as a clear evidence that further improvements and developments of relevant numerical methods would undoubtedly open up the potential of the Gerber-Shiu function widely for its practical use.

Statistical inference is an important step prior to the computation in practice \cite{Shimizu2021}.
When the surplus model is fully parameterized in advance, inference can be made relatively easily via standard estimating methodologies.
If non-trivial models, even L\'evy risk models, are chosen, one may need to employ sophisticated statistical procedures in the non/semi-parametric framework to address claims and other dispersion structures. 
To this end, the first semi-parametric framework is developed in \cite{Shimizu2011, Shimizu2012} via the regularized inversion of the Laplace transform (Section \ref{section inference - Laplace inversion}), whereas this approach is not satisfactory all the time due to the logarithmic convergence rate of the estimator \cite{Mna2008}.
Since then, a variety of statistical methods have been proposed based on the Fourier inversion \cite{Shimizu201784} (Section \ref{section inference - Fourier inversion}), and series expansions using the Fourier-sinc/cosine series \cite{wang_yu_huang_2019, wang_yu_huang_yu_fan_2019, Yang_Su_Zhang_2019, Zhang_2017} and, more intensively, the Laguerre series \cite{dussap_2021, huang_yu_pan_cui_2019, sz19, su_shi_wang_2020, su_yong_zhang_2019, su_yu_2020, zhang_su_2018, zhang_su_2019} (Section \ref{section inference - series expansion}), with which the rates of convergence are improved to polynomial orders.
In addition, statistical methods based on series expansion are of particular interest as they have the potential to reveal the asymptotic law of the estimated Gerber-Shiu function. 


The ultimate goal of the present survey is to enhance research activities on not fully explored implementation aspects of the Gerber-Shiu function by casting fresh light on its numerical and statistical methods, which seem to have been acting as a serious bottleneck in its practical use.
To this end, we begin with a comprehensive review on the formulation of the Gerber-Shiu function and its variants, the risk surplus models and additional structural features (Section \ref{section the gerber-shiu discounted penalty function}), as well as a systematic summary of analytical, semi-analytical and asymptotic methods (Section \ref{section analytical and semi-analytical approaches}).
In light of all those essential background materials, we then aim to discuss implementation aspects of the Gerber-Shiu function in two separate categories, namely, numerical methods (Section \ref{section numerical methods}) and statistical inference (Section \ref{section inference}), along with further classifications and representative concepts and formulas without going into too much nonessential technical details to avoid digressing from the main objective as a comprehensive survey.
We close the present work with concluding remarks and the future outlook on research directions and dissemination (Section \ref{section concluding remarks}) with the hope that this survey can serve as an insightful guidebook to model and method selection from practical perspectives.


%% file: sections_ver2x/edpf_section2.tex

\section{The Gerber-Shiu discounted penalty function}\label{section the gerber-shiu discounted penalty function}

Before exploring implementation aspects of the Gerber-Shiu function (Sections \ref{section numerical methods} and \ref{section inference}), we devote the present and next sections to equip ourselves with essential background materials in a systematic summary, such as its formulation, variants, the risk surplus models and additional structural features (Sections \ref{section the gerber-shiu discounted penalty function}) as well as analytical, semi-analytical and asymptotic methods (Section \ref{section analytical and semi-analytical approaches}).
Throughout the paper, we denote by $\mathbb{N}$ the set of natural numbers with $\mathbb{N}_0:=\mathbb{N}\cup \{0\}$,
and denote the Laplace transform operator associated with every $f: [0,\infty) \to \mathbb{R}$ by
\[
\mathcal{L} f(s) :=\int_0^{+\infty}e^{-su}f(u)du,
\]
for all $s$ such that the integral is finite valued. 

\subsection{Standard formulation}\label{section formulation}

We reserve the notation  
\[
U = \{U_t: \, t \geq 0\}
\] 
for the surplus process, which is a one-dimensional stochastic process with c\`adl\`ag paths almost surely (a.s.) that starts at $U_0=u$ under the probability measure $\mathbb{P}_u$, indexed by the initial state $u$.
As usual, we denote by $\mathbb{E}_u$ the expectation taken under $\mathbb{P}_u$.
Unless stated otherwise, we assume the jumps of $U$ are always downward.
In particular, if $U$ has a finite number of jumps on any non-empty interval (that is, jumps of finite activity), then it can be decomposed uniquely as
\begin{align}
U_t = V_t - L_t \label{surplus_finite_activity}
\end{align}
for a continuous process $V = \{V_t: \, t \geq 0\}$ and a pure jump process $L = \{L_t: \, t \geq 0\}$ of the form
\begin{align}
L_t := \sum_{k=1}^{N_t} Y_k, \quad t \geq 0, \label{compound_poisson}
\end{align}
with a counting process of jumps (claims) $N = \{N_t: \, t \geq 0\}$ defined by $N_t:= \# \{s\in (0,t]: L_{s} \neq L_{s-} \}$ and the corresponding jump sizes $\{Y_k\}_{k\in \mathbb{N}}$. 
For the subsequent discussions whenever $U$ admits the form \eqref{compound_poisson}, we write $\{T_k\}_{k\in \mathbb{N}}$ for a sequence of claim occurring times (so that $N_t = \max \{ k \in\mathbb{N}_0: T_k \le t \}$) and let $\{S_k\}_{k\in \mathbb{N}}$ be a sequence of interclaim times, that is, $S_k:=T_k-T_{k-1}$ for $k\in\mathbb{N}$ with $T_0:=0$.
The majority of surplus processes considered in the context of the Gerber-Shiu function is of the form \eqref{surplus_finite_activity}, although some risk processes to be reviewed in what follows have infinite number of jumps on any finite interval (that is, jumps of infinite activity) and cannot be written in terms of jump processes of the form \eqref{compound_poisson}.

The time of ruin is commonly modeled as the first down-crossing time below a certain level, most often, zero:
\begin{align}
 \tau:=\inf\left\{s\ge 0:\,U_s< 0\right\},  \label{ruin_time}
\end{align}
where it is understood that $\inf \varnothing = + \infty$. 
The Gerber-Shiu function $\phi$, introduced in \cite{Gerber_Shiu_1998}, is then defined as
\begin{equation}\label{EDPF0}
\phi(u):=
\mathbb{E}_u\left[e^{-\delta \tau}w\left(U_{\tau-},\left|U_{\tau}\right|\right)\mathbbm{1}\left(\tau<+\infty\right)\right], \quad u \geq 0,
\end{equation}
where $\delta \geq 0$ represents the constant discount rate and the so-called penalty function
\[
w: [0,+\infty)^2 \to [0,+\infty)
\] 
models the risk of the event that the surplus prior to ruin is $U_{\tau-} := \lim_{t \uparrow \tau} U_t$ and the deficit at ruin is $|U_{\tau}|$.
The indicator $\mathbbm{1}(\tau<+\infty)$ assures that  the random variables $U_{\tau-}$ and $U_{\tau}$ are well-defined on this event and also emphasizes that there is no penalization if ruin does not occur, that is, $\{ \tau = +\infty\}$. 
 
The penalty function $w$ in \eqref{EDPF0} is defined as a function of $U_{\tau-}$ and $|U_{\tau}|$, while an earlier version \cite{Gerber_Landry_1998} considers only the latter: 
with $w: [0,+\infty) \to [0,+\infty)$,
\begin{equation}\label{EDPF early version}
\mathbb{E}_u\left[e^{-\delta \tau}w\left(|U_{\tau} | \right)\mathbbm{1}\left(\tau<+\infty\right)\right]=w(0)\mathbb{E}_u\left[e^{-\delta \tau}\mathbbm{1}\left(\tau<+\infty, U_{\tau}=0\right)\right]+\mathbb{E}_u\left[e^{-\delta \tau}w\left(|U_{\tau}| \right)\mathbbm{1}\left(\tau<+\infty, U_{\tau}<0\right)\right], 
\end{equation}
where the event $\{ U_{\tau}=0 \}$ is called creeping in the literature.
Conversely, there exist a variety of generalized versions of the penalty function taking additional variables, which we summarize shortly in Section \ref{sec_generalization_w}. 
For the rest of the present survey, unless stated otherwise, the penalty function is understood to be of two variables as in \eqref{EDPF0}. 

By appropriately choosing the penalty function, various risk quantities can be modeled.
Below, we list several examples of the penalty function (see, among others, \cite{AHN2007, Asmussen_2010, cheung_liu_2022, kuznetsov_morales_2014, Landriault_Willmot_2009, ruan_yu_song_sun_huang_yu_2019, willmot_woo_2017, Zhang_2017, zhang_su_2019}): 

\noindent (1) Probability of ultimate ruin ($\delta=0$ and $w(x,y)= 1$):  
\[\mathbb{P}_u(\tau<+\infty). \]

\noindent (2) Laplace transform of the time of ruin ($\delta\ne 0$ and $w(x,y)= 1$):  
\begin{equation}\label{Sec2 Laplace transform of the time of ruin}
\mathbb{E}_u\left[e^{-\delta \tau}\mathbbm{1} (\tau< +\infty)\right].
\end{equation}

\noindent (3) Joint distribution of surplus and deficit ($\delta=0$ and $w(x,y)=\mathbbm{1}(x \le a,\,y \le b)$):  
\[
\mathbb{P}_u\left(U_{\tau-} \le a, \,|U_{\tau}| \le b, \,\tau<+\infty\right).
\]

\noindent (4) Expected discounted deficit at ruin ($w(x,y)=y$): 
\begin{equation}\label{Sec2 Expected discounted deficit at ruin}
\mathbb{E}_u \left[e^{-\delta \tau}|U_{\tau}| \mathbbm{1} (\tau< +\infty)\right].
\end{equation}

\noindent (5) Expected claim size causing ruin ($\delta=0$ and $w(x,y)=x+y$): 
\begin{equation*} 
 \mathbb{E}_u \left[|U_\tau - U_{\tau-}| \mathbbm{1}(\tau< +\infty)\right].
\end{equation*} 
			
\noindent (6) Joint moments of surplus and deficit ($\delta=0$ and $w(x,y)=x^a y^b$): 
\[ 
\mathbb{E}_u\left[U_{\tau-}^a |U_\tau  |^b\mathbbm{1} (\tau< +\infty)\right].
\] 
		
\noindent (7) An extension involving the discounted $n$th moment of the time of ruin: 
\begin{equation} \label{EDPF moments of ruin time}
\mathbb{E}_u\left[\tau^n e^{-\delta \tau}w\left(U_{\tau-},\left|U_{\tau}\right| \right)\mathbbm{1}\left(\tau<+\infty\right)\right].
\end{equation}
				
\noindent (8) Trivariate Laplace transform of the time to ruin, the surplus prior to ruin, and the deficit at ruin ($w(x,y)= \exp (-sx-zy)$): 
\begin{equation}\label{Sec2 trivariate Laplace transform}
\mathbb{E}_u \left[e^{-\delta \tau-sU_{\tau-}-z|U_{\tau}| }\mathbbm{1} (\tau< +\infty)\right].
\end{equation}  

We note that the joint Laplace transform \eqref{Sec2 trivariate Laplace transform} is deemed the most versatile for the reason that the relevant law can be fully characterized via its inversion with respect to (some of) the variables $(\delta,s,z)$.
In addition, some of the other quantities listed above can even be recovered solely through a lot more elementary operations on the Laplace transform \eqref{Sec2 trivariate Laplace transform}, such as differentiation and substitution.


%

\subsection{Generalization} \label{section Extended Gerber-Shiu function} 

While the formulation \eqref{EDPF0} covers a variety of quantities of interest, even further generalizations have often been considered, typically, in the following aspects: generalization of the penalty function (Section \ref{sec_generalization_w}), non-constant/random discounting (Section \ref{Section Generalization of the discount factor}), and finite time horizon (Section \ref{section Finite time horizon}).

%

\subsubsection{Penalty function} \label{sec_generalization_w}

In the formulation \eqref{EDPF0}, the penalty function is a function of the surplus prior to ruin $U_{\tau-}$ and the deficit at ruin $|U_{\tau}|$.
Here, we review other random elements that sometimes appear as part of the generalization of the Gerber-Shiu function. Below, we consider $U$ with downward jumps of finite activities as in the surplus process \eqref{surplus_finite_activity}. 

\begin{description}
\item[(1)] Define the total discounted claim costs until ruin:
\[
\sum_{k=1}^{N_{\tau}} e^{-\delta T_k} \kappa(Y_k)
\]
where $\kappa: (0,+\infty) \to (0,+\infty)$ quantifies the cost of each claim. With this random variable, a variant of the Gerber-Shiu function is investigated:
\begin{equation}\label{shrunk edpf}
\phi(u)=
\mathbb{E}_u\left[e^{-n_1 \delta_0 \tau}\left(\sum_{k=1}^{N_{\tau}} e^{-\delta T_k} \kappa(Y_k)\right)^{n_2} w\left(\left|U_{\tau}\right| \right)\mathbbm{1}\left(\tau<+\infty\right)\right],
\end{equation}
where 
$\delta_0\geq0$ is a constant force of interest and $n_1$ and $n_2$ are non-negative integers, with $\delta_0=\delta$ in \cite{cheung_2013}, and $n_1=1$ in \cite{liu_zhang_2015}.

\item[(2)] On the event $\{ U_\tau < 0, \tau < +\infty \}$ where $U$ downcrosses zero by a jump, it necessarily holds that $\tau = T_{N_\tau}$ and thus $U_\tau = U_{T_{N_\tau}}$.
The surplus immediately after the second to last claim before ruin $U_{T_{N_{\tau}-1}}$ is often of interest
\cite{Badescu_2009, cheung_landriault_2009, CKCheung2010, Willmot_Woo_2010}.
With this as the third argument, the resulting Gerber-Shiu function is given by
\begin{equation}\label{EDPF second last}
\phi(u)=
\mathbb{E}_u\left[e^{-\delta \tau}w\left(U_{\tau-},\left|U_{\tau}\right|, U_{T_{N_{\tau}-1}}  \right)\mathbbm{1}\left(\tau<+\infty\right)\right].
\end{equation}

\item[(3)]
Another variable that sometimes appears in the context of the Gerber-Shiu function is the last minimum of the surplus before ruin 
\begin{equation}
\underline{U}_{\tau-}:=\inf_{t\in[0,\tau)} U_t, \label{running_min}
\end{equation}
as in \cite{BIFFIS201085, kuznetsov_morales_2014}.
The corresponding Gerber-Shiu function is given by $\phi(u)=\mathbb{E}_u[e^{-\delta \tau}w(U_{\tau-},|U_{\tau}|, \underline{U}_{\tau-})\mathbbm{1}(\tau<+\infty)]$. 
Moreover, with this further introduced in \eqref{EDPF second last}, a generalized Gerber-Shiu function
\begin{equation}\label{EDPF four}
\phi(u)=
\mathbb{E}_u\left[e^{-\delta \tau}w\left(U_{\tau-},\left|U_{\tau}\right|, U_{T_{N_{\tau}-1}},  \underline{U}_{\tau-} \right)\mathbbm{1}\left(\tau<+\infty\right)\right],
\end{equation}
is investigated in \cite{CHEUNG2010}.
In a similar manner, the maximum surplus level prior to ruin $\overline{U}_{\tau-}:=\sup_{t\in[0,\tau)} U_t$ is also incorporated in \cite{CHEUNG2010127}, instead of the running minimum $\underline{U}_{\tau-}$, while \textcolor{black}{a two-sided truncated Gerber–Shiu function is investigated in \cite{woo_xu_yang_2017} in the form of
\[
\phi(u)=
\mathbb{E}_u\left[e^{-\delta \tau}w\left(U_{\tau-},\left|U_{\tau}\right|, U_{T_{N_{\tau}-1}}  \right)\mathbbm{1}\left(\overline{U}_{\tau-} < b,\, \underline{U}_{\tau-} \geq a,\, \tau<+\infty\right)\right],\quad u\in (a,b],
\]
for some $0\le a<b<+\infty$.}

\item[(4)]
\textcolor{black}{The number of claims arriving until ruin $N_{\tau}$ is incorporated into the Gerber-Shiu function in \cite{frostig_pitts_politis_2012}, as follows:
\[
\phi(u)=
\mathbb{E}_u\left[e^{-\delta \tau} z^{N_{\tau}}w\left(\left|U_{\tau}\right|\right)\mathbbm{1}\left(\tau<+\infty\right)\right],
\]
for some $|z|<1$.
By setting $\delta=0$ and $w(x)=1$, it reduces to $\mathbb{E}_u[z^{N_{\tau}}\mathbbm{1}(\tau<+\infty)]$, that is, the probability generating function of the number of claims until ruin.}

\item[(5)]
\textcolor{black}{Further to the formulation \eqref{shrunk edpf}, the joint moment of the total discounted dividends until ruin $D_{\delta}(\tau):=\int_0^{\tau}e^{-\delta t} dD_t$ 
and aggregate discounted claim costs until ruin $Z_{\delta}(\tau):=\sum_{k=1}^{N_{\tau}}e^{-\delta T_k}f(Y_k)$ 
is incorporated in \cite{cheung_liu_woo_2015}, in the form of 
\[
\phi(u)=
\mathbb{E}_u\left[e^{-\delta_1 \tau} D_{\delta_2}^n(\tau) Z_{\delta_3}^m(\tau) w\left(U_{\tau-}, \left|U_{\tau}\right|\right)\mathbbm{1}\left(\tau<+\infty\right)\right],
\quad u \in [0,b],
\]
for $m,n\in \mathbb{N}_0$, some discount rates $\delta_1,\delta_2,\delta_3$ and dividend barrier $b$.
Moreover, we refer the reader to \cite{cheung_liu_willmot_2018} for the Gerber-Shiu function with the joing moment of the total discounted gains and losses in the renewal risk model with two-sided jumps.
}


\end{description}

\subsubsection{Discount factor}\label{Section Generalization of the discount factor}

Another line of the generalization is formulated by replacing the constant discounting $\delta t$ in \eqref{EDPF0} (that is, linear in time) by a stochastic process (say, $\{\delta_t: \, t \geq 0\}$).
One such example is given in \cite{wang2017} with
$\delta_t=\delta t+\alpha M^{(1)}_t+\beta M^{(2)}_t$ for non-negative constants $\delta$, $\alpha$ and $\beta$, 
a Poisson process $\{M^{(1)}_t: \, t \geq 0\}$ and
a Gaussian process $\{M^{(2)}_t: \, t \geq 0\}$, independent of the surplus process $U$.
In a different direction, the cases where $\delta$ is dependent on the surplus $U$ have also been of interest.
For example, the so-called omega-killing model \cite{Landriault_Renaud_Zhou_2011, Li_Palmowski_2018} considers the stochastic discounting $\delta_t = \int_0^t \omega (U_s) ds$ for a non-negative measurable function $\omega$, in relation to the omega-model \cite{Gerber_Shiu_Yang_2012}.
In the Markov additive model (Section \ref{sec_Sparre-Andersen}), the discount factor is allowed to be dependent on the state of the background Markov chain \cite{Ivanovs_Palmowski_2012}.
As an alternative to the classical exponential discounting as in \eqref{EDPF0}, a version of discounting $\delta_1^{N_t}$ is considered in \cite{wang_landriault_li_2021} with respect to the number of claims $N_t$ for $\delta_1\in(0,1]$. 
%
%
%

\subsubsection{Finite time horizon}\label{section Finite time horizon}

The computation of the Gerber-Shiu function in a finite time horizon setting is often of realistic interest, as the risk management is usually planned for a short term \cite{library979190}. 
As an extension of the early version \eqref{EDPF early version}, a Gerber-Shiu type function is given by 
\begin{equation}\label{EDPF PR early}
\phi(u,T)=\mathbb{E}_u\left[w\left(U_{\tau} \right)\mathbbm{1}\left(\tau\le T\right)+P\left(U_{\tau}\right)\mathbbm{1}\left(\tau>T\right)\right],
\end{equation}
for a finite time period $T>0$ \cite{avram_usabel_2008, DIKO2011126}, where $w(\cdot)$ models the penalty in the case of ruin whereas $P(\cdot)$ models the reward for survival.
%
This is further generalized with two variables \cite{LANDRIAULT2019, Li_Lu_2017, lilusendova2019}, as in
\begin{equation}\label{finite edpf 2}
\phi(u,T)=
\mathbb{E}_{u}\left[e^{-\delta \tau}w\left(U_{\tau-},\left|U_{\tau}\right|\right) \mathbbm{1}\left(\tau\le T\right)\right], \quad \overline{\phi}(u,T)=
\mathbb{E}_{u}\left[e^{-\delta \tau}w\left(U_{\tau-},\left|U_{\tau}\right|\right) \mathbbm{1}\left(\tau >T\right)\right].
\end{equation}
The former of \eqref{finite edpf 2}, but with only $|U_{\tau}|$ in the penalty function, is discussed in \cite{xie_zhang_2021}.
Its extension with the running minimum $\underline{U}_{\tau-}$ as defined in \eqref{running_min} as the third argument is considered in \cite{kuznetsov_morales_2014}. 
An alternative formulation of the finite-time Gerber-Shiu function 
\begin{equation}\label{EDPF finite}
\phi(u,T)=
\mathbb{E}_{u}\left[e^{-\delta \tau^T}w\left(U_{\tau^T-},U_{\tau^T}\right)\right], \quad \tau^T:=\tau\land T,
\end{equation}
is studied in \cite{library979190}.  

\textcolor{black}{The expected total discounted dividends up to ruin is, strictly speaking, not a finite-time problem but can be thought of as its approximation, first discussed in \cite{cai_feng_willmot_2009_2} in the form of
$\mathbb{E}_u[\int_0^{\tau}e^{-\delta t} dD_t]$, where $D_t$ denotes the accumulated dividend paid up to time $t$.}
Subsequently, the expected discounted total cost up to default is formulated \cite{feng_shimizu_2013}, as: 
\begin{equation}\label{EDPF general total cost}
\mathbb{E}_u\left[\int_0^{\tau}e^{-\delta t} l(U_t)dt\right] = \frac{1}{\delta} \mathbb{E}_u \left[e^{- e_\delta} l (U_{e_\delta})  \mathbbm{1}\left(e_\delta < \tau \right) \right],
\end{equation}
where $l: [0,\infty) \to \mathbb{R}$ and an independent exponential random variable $e_\delta$ with rate $\delta= 1/T$ so that  the mean of $e_\delta$ is equal to $T$. 

\subsection{Surplus processes}\label{section Surplus processes}

We next provide a comprehensive review on the surplus process along with a variety of relevant extensions.
We also refer the reader to \cite{Asmussen_2010, Kyprianou2013} for details of various risk models. 

%

\subsubsection{Cram\'er-Lundberg models}\label{section Cl model}

The most classical risk process, known as the Cram\'er-Lundberg model introduced in \cite{Cramer_1930, lundberg}, is given by
\begin{equation}\label{classical process}
U_t=u+ct- L_t, \quad t \geq 0,
\end{equation}
which is within the framework \eqref{surplus_finite_activity} with $L$ being a compound Poisson process.
More precisely, the process $L$ is a special case of \eqref{compound_poisson} where $N$ is a Poisson process with rate $\lambda > 0$ (so that the interclaim times $\{S_k\}_{k\in\mathbb{N}}$ are exponentially distributed with mean $\lambda^{-1}$) and individual claim sizes $\{Y_k\}_{k\in\mathbb{N}}$ are independent and identically distributed positive random variables, independent of the interclaim times $\{S_k\}_{k\in\mathbb{N}}$.  
We note that the two sequences $\{S_k\}_{k\in\mathbb{N}}$ and $\{Y_k\}_{k\in\mathbb{N}}$ are sometimes assumed dependent \cite{Boudreault2007, xie2017}. 

Regarding the claim size $Y$, the most common choices (because they often lead to analytical expressions) are exponential \cite{Albrecher_Hartinger_2007, LABBE2011, Li_Garrido_2004, liu_zhang_2015, xie2017}, hyper-exponential \cite{YTChen2007, Lin_Willmot_1999, WANG2008} and Erlang distributions \cite{ALBRECHER2010, Willmot_Woo_2010}, which are in the class of phase-type distributions (Section \ref{sec_Sparre-Andersen}).
Note that analytical expressions are not available for general claim size $Y$, whereas renewal theoretic arguments often provide asymptotic results (Section \ref{subsubsection Asymptotic results}) that differ depending on whether the claim size distribution is light- or heavy-tailed \cite{Asmussen_2010}. 


\subsubsection{L\'evy risk model}\label{section Levy risk model}


The L\'evy process is a continuous-time stochastic process with independent and stationary increments with c\`adl\`ag paths. 
The class of L\'evy processes is a lot richer than the Cram\'er-Lundberg model \eqref{classical process} in the sense that it can accommodate jumps of infinite activity and/or paths of unbounded variation.

In our context, a convenient and concise characterization of the L\'evy process can be given in terms of the Laplace exponent.
For any L\'evy process $X = \{X_t: \, t \geq 0\}$, we denote its Laplace exponent by
\begin{align}
\Psi_X(\theta) := \log \mathbb{E} [e^{\theta X_1}], \label{Psi_X}
\end{align}
for all $\theta$ such that the expression is finite valued.
The characteristic exponent (with $\theta$ replaced with $i \theta$) always exists and is thus often preferred for the study of general L\'evy processes. 
However, for the study of the Gerber-Shiu function, which usually considers the surplus process with only downward jumps (spectrally negative L\'evy processes), it is sufficient in many instances to use the Laplace exponent since then it is well defined for every $\theta \geq 0$ and uniquely characterizes the marginal law just as the characteristic exponent does.
As discussed later in Section \ref{section closed - Scale functions}, the so-called scale function plays an important role in the recent study of the Gerber-Shiu function on spectrally negative L\'evy processes.

When the surplus process $U$ is a spectrally negative L\'evy process with a finite first moment, it follows from the L\'evy-Khintchine formula that 
\begin{equation}
\Psi_U(\theta)= \log \mathbb{E} [e^{\theta U_1}] = \theta a+\frac{1}{2}\sigma^2\theta^2+\int_{0+}^{+\infty}\left(e^{-\theta x} - 1 + \theta x\right)\Pi(dx), \quad \theta \geq 0, \label{laplace_exp_perturb}
\end{equation}
where $a \in \mathbb{R}$, $\sigma\geq 0$, and the L\'evy measure $\Pi$ (of the dual process $-U$)  satisfies the integrability condition $\int_{(0,+\infty)}(1 \wedge x^2)\Pi(dx)<+\infty$.
We note that the term ``$\theta x$'' in the integrand is usually cut off, like ``$\theta x \mathbbm{1}(x < 1)$,'' when the marginal $U_1$ does not have a finite first moment (that is, $\mathbb{E} [U_1] = -\infty$). 
Also, the first moment $\mathbb{E}[U_1]$ is often assumed to be strictly positive in the relevant literature, since then  the surplus diverges ($\lim_{t \to +\infty}U_t = +\infty$, $a.s.$) and thus ruin does not always occur ($\mathbb{P} (\tau < + \infty) < 1$). 

A spectrally negative L\'evy process $U$ is of bounded variation if and only if $\sigma = 0$ and $\int_{0+}^1 x\Pi(dx)<+\infty$, with the exponent \eqref{laplace_exp_perturb} reduces to
\[
\Psi_U(\theta)= c\theta +\int_{0+}^{+\infty}\left(e^{-\theta x} - 1 \right)\Pi(dx), \quad \theta \geq 0,
\]
with $c:= a + \int_{0+}^{+\infty} x \Pi(dx)$.
If, moreover, $\Pi(0,+\infty)<+\infty$, then it is of finite activity and admits the form \eqref{surplus_finite_activity}; otherwise ($\Pi(0,+\infty)=+\infty$), it is said to be of infinite activity.

A L\'evy process $\{L_t:\,t\ge 0\}$ is called a subordinator if its paths are almost surely non-decreasing.
With no drift term, its Laplace exponent is then given by 
\begin{align} \label{Laplace_subordinator}
\Psi_L(\theta) = \log \mathbb{E} [e^{\theta L_1}] = \int_{0+}^{+\infty} \left(e^{\theta x}-1\right) \Pi(dx),
\end{align}
which is well defined at least for all $\theta \le 0$.
In particular, if $\lambda = \Pi(0,+\infty)<+\infty$, a subordinator reduces to a compound Poisson process, like in the Cram\'er-Lundberg model \eqref{classical process}.
In this case, with $F_Y$ the distribution of the claim size $Y$,
one can write $\Pi(dx) = \lambda F_Y(dx)$ on $(0,+\infty)$ and thus $\Psi_L(\theta) = \lambda \int_{0+}^{+\infty} (e^{\theta x}-1) F_Y(dx) = \lambda (\mathbb{E} [e^{\theta Y_1}] -1)$.
Moreover, 
the Laplace exponent of the surplus process $U$ in the Cram\'er-Lundberg model \eqref{classical process} can then be written as
\begin{align}
\Psi_U(\theta) = \log \mathbb{E} \left[e^{\theta U_1}\right] =  c \theta  + \Psi_L(- \theta),  \label{Laplace_CR}
\end{align}
which is well defined at least for all $\theta \geq 0$.
It is worth mentioning that the negative of a subordinator is usually not categorized in the class of spectrally negative L\'evy processes.

\subsubsection{Diffusion perturbation} \label{section_diffusion_perturbation}

As a generalization of the Cram\'er-Lundberg model \eqref{classical process}, its diffusion perturbation has long attracted a great deal of interest in modeling small perturbations caused by measurement errors, deviation of the premiums and claims \cite{Asmussen_2010}, given by
\[
U_t=u+ct- L_t + \sigma W_t, \quad t \geq 0, 
\]
with a standard Wiener process $W = \{W_t: \, t \geq 0\}$
(see, among others, \cite{avram_usabel_2008, cheung_liu_2022, lilusendova2019, Li_Wu_2006, lutsai2007,  TSAI200251}).
This generalization carries over the diffusion component of \eqref{laplace_exp_perturb} on top of the expression \eqref{Laplace_CR}, with its Laplace exponent given by $\Psi_U(\theta) =  c \theta + \frac 1 2 \sigma^2 \theta^2 + \Psi_L(-\theta)$ for $\theta \geq 0$.

\subsubsection{Sparre-Andersen models} \label{sec_Sparre-Andersen}

The Cram\'er-Lundberg model \eqref{classical process} has also been generalized by employing a more general claim arrival process, rather than using an ordinary Poisson process.
One important example in this direction is the Sparre-Andersen risk model \cite{Andersen_1957}, given by a modification of \eqref{classical process}, with the Poisson process $\{N_t:\,t\ge 0\}$ replaced by a suitable renewal process, whose interclaim times $\{S_k\}_{k\in\mathbb{N}}$ form a sequence of iid positive random variables, rather than iid exponential random variables.
Again, the inter-claim times $\{S_k\}_{k\in\mathbb{N}}$ and claim sizes $\{Y_k\}_{k\in\mathbb{N}}$ are usually assumed independent, with some exceptions 
 \cite{Badescu_2009, ORBANMIHALYKO2011}. 

The Sparre-Andersen model is in general not analytically tractable, whereas if the interclaim times are of phase-type, then semi-analytical results are often available. 
A phase-type distribution is concerned with the first absorption time of a finite-state continuous-time Markov chain on a state space, consisting of a finite number of transient states $E:=\{1,2,\cdots, n\}$ and a single absorbing state. 
Its initial distribution on the state space $E$ is given by an $n$-dimensional row vector ${\bf \alpha} = (\alpha_1, \cdots, \alpha_n)$, while the transition matrix is given by
\[
\left( \begin{array}{ll}  {\bf T} & {\bf t}  \\  0^\top & 0 \end{array} \right),
\]
where ${\bf t}:=- {\bf T} {\bf 1}$ with the column vector of ones $\bf 1$.
The exponential distribution with rate $\lambda > 0$ is its special case with $E= \{1 \}$, ${\bf T} = -\lambda$ and ${\bf t} = \lambda$. 
The Erlang distribution with shape parameter $n$ can be written as a phase-type distribution with $E= \{1,\cdots, n \}$ with 
\[
{\bf T}_{ij} = \left\{ \begin{array}{ll} -\lambda, & \text{if } i = j, \\ \lambda, & \text{if }j = i+1, \\ 0, & \textrm{otherwise,} \end{array} \right. \quad \textrm{and} \quad {\bf t}_i = \left\{ \begin{array}{ll} 0, & \text{if }i \neq n, \\ \lambda, & \textrm{otherwise.} \end{array} \right. 
\]
The hyper-exponential distribution, or a mixture of exponential distributions with rate $\lambda_i$ for $i\in \{1,\cdots,n\}$, is also of phase-type with
\[
{\bf T}_{ij} = \left\{ \begin{array}{ll} -\lambda_i & \text{if }i = j \\ 0 & \textrm{otherwise} \end{array} \right. \quad \textrm{and} \quad {\bf t}_i = \lambda_i.
\]
The Coxian distribution is another important class of the phase-type distribution, modeled by a Markov chain with transition matrix of the
form
\[
{\bf T}_{ij} = \left\{ \begin{array}{ll} -\lambda_i, & \text{if } i =
j, \\ p_i \lambda_i, & \text{if }j = i+1, \\ 0, & \textrm{otherwise,}
\end{array} \right. \quad \textrm{and} \quad {\bf t}_i = \left\{
\begin{array}{ll} (1-p_i) \lambda_i, & \text{if }i \neq n, \\
\lambda_i, & \textrm{otherwise.} \end{array} \right.
\]
It reduces to the (generalized) Erlang and hyperexponential distribution when $p_i \equiv 1$ and $p_i \equiv 0$, respectively.
It is known to be almost as general as the phase-type distribution itself, in the sense that every phase-type distribution that can be represented with acyclic Markov chains has an equivalent Coxian representation.
A variety of the Sparre-Andersen models in the literature are built with phase-type distributed interclaim times.
For example, the Erlang(n) case has been studied in \cite{Gerber_Shiu_2005, Li_Garrido_2005,  zhang_li_yang_2009}. 

As an extension, in the stationary Sparre-Andersen model, claims arrive according to a stationary renewal process $N$,
so that the law of the process $\{N_{t+s} - N_t:\, s \ge 0\}$ is invariant in $t$
\cite[Section V.3]{Asmussen_2003}. 
The Gerber-Shiu function on the stationary Sparre-Andersen model can often be written in terms of that on the ordinary Sparre-Andersen model \cite{WILLMOT2003}. 
Another stationary Sparre-Andersen model can be found in \cite{PAVLOVA2004} in the discrete 
setting.
Other variants of the Sparre-Andersen model include 
a delayed Sparre-Andersen model \cite{willmot_2004} and its extensions where the premium $ct$ in \eqref{classical process} is replaced by an independent Poisson process 
\cite{bao_ye_2007} and with a multi-layer dividend strategy \cite{DENG2012}.
\textcolor{black}{We refer the reader to the monograph \cite{willmot_woo_2017} for more detail on the Sparre-Andersen model in the context of the Gerber-Shiu function.}

\subsubsection{Markov additive processes} \label{subsubsection markov additive processes}

Renewal processes with phase-type interclaim times are often mathematically tractable owing to its equivalent representation as a Markov-modulated Poisson process \cite{Asmussen_2003}.
To be more precise, a renewal process can be represented as a Markovian arrival process whose background continuous-time Markov chain is a variant of the one appeared in Section \ref{sec_Sparre-Andersen}, here with transition matrix ${\bf T} + {\bf t \bf \alpha}$ where ${\bf T}$ describes the state transition without arrivals and ${\bf t \bf \alpha}$ with arrivals.
By generalizing Markov-modulated Poisson processes, the Sparre-Andersen process can be written as a Markov additive process (MAP), which is a bivariate Markov process $(U,J)$ where the increments of the surplus process $U$ are governed by the continuous-time Markov chain $J$ in a sense.
Conditionally on the state of the Markov chain (say, $J = i \in \mathcal{E}$), the surplus process $U$ evolves as a L\'evy process (say, $U^{(i)}$), until the Markov chain makes a transition to another state ($j \in \mathcal{E}$), at which instant an independent jump specified by the pair $(i,j)$ is introduced into the surplus process $U$.
In the case of the Sparre-Andersen process in the form \eqref{classical process} with phase-type interclaim times, the corresponding MAP $(U,J)$ can be characterized by the transform
\[
\mathbb{E} \left[e^{\theta (U_t-U_0)} \mathbbm{1}(J_t = j) \Big|\, J_0 = i \right]  = \left(e^{{\bf F}(\theta)t}\right)_{i,j}, \quad {\bf F}(\theta):=c \theta \mathbb{I}_n+ {\bf T}+ \mathbb{E} \left[e^{-\theta Y_1}\right]  {\bf t \alpha}, 
\]
for $i,j \in \mathcal{E}$, $t \geq 0$ and $\theta \ge 0$, 
where $\mathbb{I}_n$ denotes the identity matrix of order $n$.
MAPs are first introduced in \cite{Asmussen_1989} to model surplus in risk theory. 
The MAP risk model is flexible enough to handle cases where the claim size as well as the discount factor depend on the state transition (that is, $i$ and $j$) \cite{Ivanovs_Palmowski_2012}.
The Gerber-Shiu function on Markov-modulated Poisson processes has been studied in \cite{Asmussen_2010, lilu2008, lutsai2007}.
A variant of the Gerber-Shiu function incorporating the maximum surplus prior to ruin in a MAP risk model is investigated in \cite{CHEUNG2010127}.
\textcolor{black}{A discrete-time counterpart, the so-called Markov additive chain, has recently been introduced in \cite{palmowski_ramsden_papaioannou_2022} for addressing the discrete-time Gerber-Shiu function.}

\textcolor{black}{We remark that  the surplus process can also be analyzed by embedding a fluid queue process \cite{Asmussen_1995, Ramaswami_2006}, based on which the Gerber-Shiu function can be computed when the claims occur according to a Markovian arrival process and the claim size is phase-type distributed \cite{AHN2007}.}

\subsubsection{Surplus processes with additional components}\label{subsubsection surplus processes with additional components}

The Cram\'er-Lundberg model \eqref{classical process} has been generalized by superimposing a stochastic process $\{Z_t:\,t\geq 0\}$ onto the surplus process $\{U_t:\,t\ge 0\}$, like $U_t=u+ct- L_t + Z_t$, for instance, for modelling additional random income or claims.
In fact, the diffusion perturbation (Section \ref{section_diffusion_perturbation}) can also be interpreted in this way with $Z = \sigma W$.
Other examples in the literature include Poisson  \cite{bao_ye_2007, BAO2006} and compound Poisson processes  \cite{CHADJICONSTANTINIDIS2009, gao_wu_2014, LABBE2009, ZHAO2012}.
In the presence of upward jumps in the additional component, the resulting surplus process is no longer spectrally negative, while it can still be written as a Markov additive process (Section \ref{subsubsection markov additive processes}) as long as the positive jump size distribution is of phase-type \cite{Asmussen_Avram_Pistorius_2004}. 
The risk model containing by-claims induced by main claims can be modeled by setting this additional component to the negative of the running sum of by-claims \cite{gao_wu_2014, XIE2011, zou_xie_2012}.

\subsection{Specialized formulations for applications}\label{subsection formulations specialized for applications}


We close the present section with several practical applications owing to specialized formulations of the surplus process, the penalty function and more generally the Gerber-Shiu function per se, such as tax, dividends and interests (Section \ref{Drift component}), credit default swaps (Section \ref{section CDS}), capital injection (Section \ref{subsubsection capital injection}) and risk measures (Section \ref{subsubsection risk measures}). 

\subsubsection{Tax, dividends and interests}\label{Drift component}



In the Cram\'er-Lundberg framework, if the surplus earns interest at a constant rate $r$, then the risk process can be described \cite{cai_feng_willmot_2009, SCHMIDLI2015, WANG2008, wulufang2007} as
\begin{equation}\label{constant interest process}
U_t=ue^{rt}+c\int_0^t e^{rs}ds-\int_0^t e^{r(t-s)}d L_s.
\end{equation}
If the insurer is allowed to borrow at a constant force of debit interest $r_2$ when the surplus becomes negative while interest is earned at a constant rate $r_1$ otherwise,
then the risk process can be written in the form of stochastic differential equation \cite{LI2013}:
\begin{equation} \label{eq_LI2013}
dU_t=\left(c+\left(r_1 \mathbbm{1}\left(U_t\geq 0\right)+r_2 \mathbbm{1}\left(U_t< 0\right)\right)U_t\right)dt-dL_t.
\end{equation}
The constant force of interest $r$ is replaced by a suitable stochastic rate in \cite{avram_usabel_2008, DIKO2011126, Yuen_Wang_2005}. 
Tax payment at a fixed rate $\gamma$ is incorporated \cite{Ming_Wang_Xiao_2010} together with a constant debit interest, where the surplus process is modelled as
\begin{equation} \label{eq_ming_wang}
dU_t=\left(c-c\gamma\mathbbm{1}\left(U_t= 
\sup\nolimits_{s\in[0,t]}U_s,
\, U_t\geq 0\right)+r_2 \mathbbm{1}(U_t<0)U_t\right)dt-dL_t.
\end{equation}
A constant dividend strategy \cite{CHADJICONSTANTINIDIS2009, CHEUNG2010127, Gerber_Shiu_Yang_2010, lilu2008, Li_Wu_2006,  Lin_Willmot_Drekic_2003, yuen_wang_li_2007}, which is only slightly outside the realm of the Cram\'er-Lundberg framework \eqref{classical process}, 
is formulated via the surplus process 
\begin{equation}\label{constant dividend surplus process}
dU_t=c\mathbbm{1}\left(U_t<b\right)dt-dL_t,
\end{equation}
where any excess of the surplus above a certain level $b$ is to be paid as dividend.
In such dividend strategies, the present value of the total dividends paid up to ruin can be expressed as \eqref{EDPF general total cost}, where $l(U_t)$ here represents the dividend rate depending on the surplus $U_t$ \cite{Cai_Gerber_Yang_2006, feng_shimizu_2013, Gerber-Shiu_2006}.
More generally, 
a reflected process $U_t = X_t - \sup_{s\in [0,t]} (X_s - b)$ with a general L\'evy process $X$ has attracted attention in the context of de Finetti's optimal dividend problem, which is concerned with 
the optimality of such barrier dividend strategies.
As another generalization, reflected MAPs have also been investigated \cite{Ivanovs_Palmowski_2012}.



The surplus process \eqref{constant dividend surplus process} has been further generalized with layers by replacing the constant barrier $b$ with a linear barrier $b(t)=b_0+at$ \cite{Albrecher2005}, where the dynamics is given by $dU_t=(c\mathbbm{1}(U_t<b_0+at)+a\mathbbm{1}(U_t\geq b_0+at))dt-dL_t$.
Other dividend strategies include impulsive dividend strategies \cite{LIU2015}, resulting in a lower level of the  surplus $a<b$ through a dividend payment when it reaches the constant barrier level $b$, and the multi-threshold dividend strategy \cite{Albrecher_Hartinger_2007, DENG2012, Lin_Sendova_2008,  xie2017, YANG2008, zhang_zhou_guo_2006}, which models the premium rate by a step function of the risk surplus in order to justify the ratio of earnings retention and dividend payments. 
In the latter strategy, the surplus process is given by $dU_t=\sum_{k=1}^M c_k \mathbbm{1}(U_t\in [b_{k-1},b_k))dt-dL_t$, with threshold layers $0=b_0<b_1<\cdots<b_{M-1}<b_M=+\infty$.
A variant driven by a L\'evy process is studied in \cite{czarna_2019} 
as a generalization of the refracted L\'evy process \cite{Kyprianou_Loeffen_2010}.
A related three-barrier model is investigated recently in \cite{zhang2022risk}.

\textcolor{black}{We add that in the Sparre-Andersen framework (Section \ref{sec_Sparre-Andersen}), those formulations have been studied extensively in \cite{cheung_2011} along with a variety of examples for the generalized Gerber-Shiu function \eqref{EDPF second last} where the premium rate depends on the surplus subject to threshold dividend strategy, credit or debit interests, and the absolute ruin problem.} 

\subsubsection{Credit default swaps}\label{section CDS}



The Gerber-Shiu function is closely related to  
the structural approach of credit risk analysis, which is a major subfield of mathematical and corporate finance,
where the default of a firm is defined as the first down-crossing time of the asset price dynamics \cite{Schoutens_Cariboni_2009}, in a similar manner to how the ruin time $\tau$ is defined in \eqref{ruin_time}. 
In particular, the fair premium of a credit default swap (CDS) contract can be expressed in terms of a combination of finite-time Gerber-Shiu functions \cite{Hao_Li_Shimizu_2013} in a simple case where the recovery rate is a constant, and is further generalized as well when the recovery rate depends on the firm value at bankruptcy. 

Consider a defaultable bond with maturity $T$ of a firm whose asset value dynamics is given by $V_t = V_0 \exp(X_t)$.
The driving process $X_t (= \log (V_t/V_0))$ in the exponent is often set to a Brownian motion or a L\'evy process in the relevant literature.
We assume that the firm defaults at the moment when the asset value goes below a certain level, say $d (< V_0)$. 
By setting $U_t=  u + X_t$ with $u= \log (V_0/d)$, we have that $V_t < d$ if and only if 
$U_t <0$, and hence the time of default can be written exactly the same as the ruin time \eqref{ruin_time} in the context of the Gerber-Shiu function.
That is, the structural model of credit risk has now been reformulated into the Gerber-Shiu framework by interpreting the log asset value as the surplus process.

The main objective of the credit risk research is to evaluate the fair premium, so that the expected cash flow between the protection buyer and seller is zero.
This quantity of interest depends on the recovery rate, that is often assumed to be a function of the asset value upon bankruptcy, say $R(U_{\tau})$. 
A protection buyer, aiming to hedge default risk,
buys a CDS with constant premium rate $P\in (0,1)$ to receive the $1-R(U_{\tau})$ fraction of the face value (called default payment) upon default, if it occurs.  
The counterpart, a protection seller, on the other hand, pays the default payment when default occurs while receiving the premium up to maturity $T$ or default, whichever comes first.
To be fair to both parties, the premium $P$ should be set in such a way that the net present values of the expected payoffs coincide, that is,
\[
\mathbb{E}\left[\int_0^{\tau\land T} e^{-\delta s}P \,ds \right] = \mathbb{E}\left[e^{-\delta \tau}(1 - R(U_\tau)) \mathbbm{1}(\tau \le T)\right], 
\]
with a constant interest rate $\delta>0$. 
By solving this equation for $P$, one finds the fair premium rate $P^*$ in terms of two types of finite-time Gerber-Shiu functions \eqref{finite edpf 2} and \eqref{EDPF finite}: 
\[
P^* = \frac{\mathbb{E}[e^{-\delta \tau}(1 - R(U_\tau)) \mathbbm{1}(\tau \le T)]}{\delta^{-1}\mathbb{E}[1 - e^{-\delta (\tau\land T)}]}. 
\]

%
It is worth noting that the study of the Gerber-Shiu function has also been very close to other types of research in finance.
For example, 
a risk measure is investigated in \cite{shimizu_tanaka_2018} in the framework of the Gerber-Shiu function for the evaluation of solvency and default risks in a variety of financial contexts.

\subsubsection{Capital injection}\label{subsubsection capital injection}


Another practical application of the Gerber-Shiu function can be found in the study involving capital injections, which aims to 
maintain the solvency of the portfolio \textcolor{black}{\cite{Eisenberg_Schmidli_2011, nie_dickson_li_2011, Schmidli_2014, yu_guo_wang_guan_huang_yu_2021}}.
Given a surplus process $U$ and $\tau$, the first time $U$ goes below zero (which corresponds to the time of ruin in the Gerber-Shiu framework), the capital injection model is designed to inject the capital $|U_\tau|$ (or more), taken out from an external source, to the company in order to avoid its bankruptcy.
By repeatedly receiving capital injection, the company avoids ruin permanently.


Let $Z=\{Z_t\,:\,t\ge 0\}$ be a process that aggregates capital injection in such a way that bankruptcy never occurs, that is, $U_t + Z_t \ge 0$, $a.s.$ 
The primary objective here is to minimize the net present value of the total capital injection $\mathbb{E}_u[\int_0^{+\infty} e^{-\delta t} dZ_t]$ over all injection strategy $Z$ under the condition $U_t + Z_t \ge 0$, $a.s.$, that is,
$f^*(u):= \inf_Z \mathbb{E}_u[\int_0^{+\infty} e^{-\delta t} dZ_t]$.
The minimizer should first inject exactly $|U_\tau|$ to bring the surplus back up to zero, so that the jump size $dZ_t$ then can be kept as small as possible. 
If the surplus process is time-homogeneous and strongly Markovian, then it holds by the renewal argument that the minimum amount of injection after the first ruin is $f^*(0)$, and thus moreover, 
\[
f^*(u) = \mathbb{E}_u\left[e^{-\delta \tau}\left( f^*(0) + |U_\tau|\right) \mathbbm{1}(\tau < +\infty)\right], 
\]
which is nothing but a combination of the two formulations \eqref{Sec2 Laplace transform of the time of ruin} and \eqref{Sec2 Expected discounted deficit at ruin} of the Gerber-Shiu function
\cite{Schmidli_2014}.  

\textcolor{black}{Closely related to capital injection is reinsurance. 
For instance, in \cite{nie_dickson_li_2011}, a capital injection strategy with positive barrier $k$ is formulated as a minimization problem of the probability of ultimate ruin, modeled by the event that the deficit exceeds $k$ (and hence the pre-capital-injection surplus downcrosses zero). 
This formulation is considered as a type of reinsurance contract that restores the insurer's surplus to a level $k$.
In a more standard reinsurance policy problem, the objective is set to select an optimal reinsurance contract, or a mapping from the claim size $Y_k$ to the portion the first insurer needs to cover.
The drift term there is modified by the reinsurance premium which depends on the contract.
Also, in \cite{preischl_thonhauser_2019}, a problem of minimizing the Gerber-Shiu function is addressed over a set of reinsurance strategies in the Cram\'er-Lundberg model.}


\subsubsection{Risk measures}\label{subsubsection risk measures}

The Gerber-Shiu function has also attracted attention in the context of ruin-related risk measures, naturally because bankruptcy is the risk that insurance companies wish to avoid most.
For instance, risk measures are developed based on ruin probabilities and applied to the construction of reinsurance strategies, such as a VaR-type risk measure \cite{Trufin_Albrecher_Denuit_2011} on the Cram\'er-Lundberg model \eqref{classical process}: for $\epsilon \in (0,1)$, 
\begin{equation}\label{operator rho}
\rho_\epsilon[Y] := \inf_{u\in (0,+\infty)}\left\{\mathbb{P}_u(\tau < +\infty) \le \epsilon\right\} = \inf_{u\in (0,+\infty)}\left\{\mathbb{P}\left(\sup\nolimits_{t \in [0,+\infty)} \left\{L_t - ct\right\}>u\right) >1 - \epsilon \right\}, 
\end{equation}
with respect to the law of the claim size $Y$, which returns the minimum initial reserve that keeps the ruin probability less than $\epsilon$. 
It is known \cite{Trufin_Albrecher_Denuit_2011} that the map \eqref{operator rho} satisfies the definition of a risk measure in the financial context \cite{Artzner_1999}.
This approach has been tailored to the Gerber-Shiu framework \cite{Eisenberg_Schmidli_2011, Schmidli_2002, Schmidli_2014} for measuring the minimum initial reserve that controls ``Gerber-Shiu type'' risks, such as total capital injections (Section \ref{subsubsection capital injection}), whereas those developments do not meet all requirements to be a risk measure in the sense of \cite{Artzner_1999}, that is, it is not entirely clear what the ``risk'' measured (that is, the law of the claim size $Y$) represents. 
Most recently, now in line with the framework \cite{Artzner_1999}, a dynamic version is developed \cite{shimizu_tanaka_2018} based on the finite-time Gerber-Shiu function (Section \ref{section Finite time horizon}) in the form of the so-called Gerber-Shiu process $\{\phi_t(u,T):\,t\in [0,T]\}$, defined by 
\begin{equation}\label{def of gerber-shiu process}
\phi_t(u,T) := \begin{cases}
\mathbb{E}_u[e^{-\delta (\tau\wedge T)} w(X_{(\tau\land T)-}, |X_{\tau\wedge T}|)|\,\mathcal{F}_t],& \text{on } \{\tau > t\}, \\
+\infty, & \text{on } \{\tau \le t\},
\end{cases} 
\end{equation}
where $\{X_t:\,t\in [0,T]\}$ denotes a general surplus process with initial state $X_0=u$ and its natural filtration $(\mathcal{F}_t)_{t\in [0,T]}$.
The Gerber-Shiu risk measure is then proposed with respect to time, as follows: 
\[
{\rm GS}_{t,T}^\epsilon(X):= \inf_{z\in \mathbb{R}}\left\{\phi_t(u + z,T) < \epsilon\right\}, 
\]
which represents the minimum amount to be adjusted at time $t$ to keep the Gerber-Shiu process less than $\epsilon$ all the way up to maturity $T$.
That is, this map measures a dynamic risk in time on the path space of the trajectory $X$ (for instance, the Skorokhod space $\mathbb{D}$ of c\`adl\`ag functions).
Despite this dynamic risk measure is rich enough to accommodate various practical problems, the computation of the conditional expectation involved in the Gerber-Shiu process \eqref{def of gerber-shiu process} is not a trivial task as of yet.   
On the whole, the risk measure in the Gerber-Shiu framework would be an important future line of research deserving of separate investigation.

%% file: sections_ver2x/edpf_section3.tex

\section{Analytical, semi-analytical and asymptotic methods}\label{section analytical and semi-analytical approaches}

In the literature, the evaluation of the Gerber-Shiu function has been tackled quite extensively via a wide variety of analytical, semi-analytical and asymptotic approaches under restrictive conditions imposed in terms of formulation and underlying surplus processes (Section \ref{section the gerber-shiu discounted penalty function}), thus often without strict need for numerical approximation, yet at a great sacrifice of richness and versatility of the Gerber-Shiu function.
The primary objective of the present section is to complete relevant knowledge ready on the table prior to the main sections on implementation aspects (Sections \ref{section numerical methods} and \ref{section inference}) by providing a comprehensive survey on the existing theoretical frameworks leading to analytical, semi-analytical and asymptotic expressions.

\subsection{Laplace transform of the Gerber-Shiu function}\label{section closed - Laplace transform}

The Laplace transform of the Gerber-Shiu function is often easier to find than the exact expression for the Gerber-Shiu function itself.
We review the existing Laplace transform based approaches when the risk process $U$ is a L\'evy process and other related stochastic processes.
When $U$ is a L\'evy process, the so-called (Cram\'er-)Lundberg equation is given by
\begin{equation}\label{Lundberg equation}
	\Psi_U(x)  = \delta,
\end{equation}
where $\delta\ge 0$ and $\Psi_U$ is the Laplace exponent defined in Section \ref{section Levy risk model}.
The largest non-negative solution $\Phi(\delta)$ to the Lundberg equation (that is, $\Psi_U(\Phi(\delta)) = \delta$) is an important quantity, whose role 
will be made clear in what follows.
\textcolor{black}{We remark that (negative) solutions to the Lundberg equation \eqref{Lundberg equation} are often assumed to be all distinct, as opposed to the fact that there may exist a (negative) solution of multiplicity greater than one.
This simple-root assumption is not restrictive in practice, that is, when the Laplace exponent $\Psi_U$ is a rational function, since then there can exist only a finite number of non-negative constants $\delta$ with which the Lundberg equation \eqref{Lundberg equation} has a solution of multiplicity greater than one \cite[Proposition 5.4]{kuznetsov_2013}. 
For more detail on such exceptions, we refer the reader to, for instance, \cite{Asmussen_Avram_Pistorius_2004, LABBE2011}.}


\subsubsection{Cram\'er-Lundberg and L\'evy models}

The Laplace transform is closely related to the (integro-)differential equation, which can be derived in various manners.
One approach is to use the fact that the discounted and killed process $\{e^{-\delta (t \wedge \tau)} \phi (U_{t \wedge \tau}):\, t \ge 0\}$ is a local martingale. 
If the Gerber-Shiu function $\phi$ is sufficiently smooth for the Ito formula, the Lundberg equation \eqref{Lundberg equation} yields the identity $\mathcal{A} \phi(u) = \delta \phi(u)$, where $\mathcal{A}$ is the infinitesimal generator of the L\'evy process $U$. 
This approach relies on the differentiability of the Gerber-Shiu function $\phi$ (typically depending on the path variation), which holds true in many instances \cite[Chapter XII.2]{Asmussen_2010}.

As for the Cram\'er-Lundberg model \eqref{classical process}, a more intuitive derivation is available.
By conditioning on the time and amount of the first claim before time $h(> 0)$ (provided that there is such a claim), we obtain 
\begin{align}\label{Laplace 1}
	\phi(u)=e^{-(\delta+\lambda)h} \phi(u+ch)+\int_0^{h}\lambda e^{-(\lambda+\delta) t} \left[\int_0^{u+ct}\phi(u+ct-x)p_Y(x)dx + \int_{u+ct}^{+\infty}w(u+ct, x-u-ct)p_Y(x)dx\right]dt,
\end{align}
where $c$ and $\lambda$ denote the constant premium rate and the unit-time intensity of the homogeneous Poisson process in \eqref{classical process}. 
Hereafter, we assume that the claim size $Y$ admits a density $p_Y$ so that $F_Y(dx) = p_Y(x) dx$ on $(0,+\infty)$ for convenience.
By differentiating \eqref{Laplace 1} with respect to $h$ at $0+$, we obtain 
\begin{equation}\label{Laplace 2}
	-(\lambda+\delta)\phi(u)+c\phi'(u) + \lambda\int_0^u\phi(u-x)p_Y(x)dx+\lambda\int_u^{+\infty}w(u,x-u)p_Y(x)dx=0.
\end{equation}
Taking the Laplace transform on both sides of the integro-differential equation \eqref{Laplace 2} yields the Laplace transform of the Gerber-Shiu function $\mathcal{L}\phi(s)=\int_0^{+\infty}e^{-su}\phi(u)du$. 
%
In particular, we obtain, for $s>\Phi(\delta)$,
\[
\mathcal{L} \phi(s) 
=\frac{\lambda\left( \mathcal{L} \omega(\Phi(\delta))- \mathcal{L} \omega(s) \right)}{cs+\lambda (\mathbb{E}[e^{-s Y_1}]-1)-\delta},
\]
where $\omega(s) :=\int_0^{+\infty}e^{-s u}\int_u^{+\infty}w(u,x-u)p_Y(x)dxdu$.
Here, 
we recall that $\Phi(\delta)$ is the unique non-negative solution to the Lundberg equation \eqref{Lundberg equation}, or more precisely in this case, 
\begin{equation}\label{Lundberg equation 1}
	\lambda\left(\mathbb{E}\left[e^{-\Phi(\delta) Y_1}\right]-1\right)+c \Phi(\delta) -\delta=0.
\end{equation}
We note that the Dickson-Hipp operator \cite{Dickson_Hipp_2001, Li_Garrido_2004} has often been involved in the computation of the Laplace transform of the Gerber-Shiu function. 




After its first development \cite{Gerber_Shiu_1998} for the Cram\'er-Lundberg model \eqref{classical process}, the Laplace transform based method has been employed, extended and improved in a wide variety of problem settings.
Some examples include
dependence between interclaim arrivals and claim sizes \cite{Boudreault2007, ORBANMIHALYKO2011}  together with perturbation \cite{Zhang_Yang_2011},  perturbation with two-sided jumps \cite{Zhang_Yang_Li_2010}, perturbation and investment with penalty and reward in a finite time \cite{avram_usabel_2008}, delayed claims induced by main claims \cite{XIE2011, zou_xie_2012}, random income modeled by a compound Poisson process \cite{Albrecher_Gerber_Yang_2010} and additionally with delayed-claims  \cite{gao_wu_2014, zhu_huang_yang_zhou_2014}, constant interests that can be positive or negative \cite{SCHMIDLI2015}, capital injection restoring the surplus to a certain level \cite{Dickson2016}, a finite-time problem with and without perturbation \cite{Li_Lu_2017, lilusendova2019}, 
and a generalized stochastic income with additional discounting  \cite{wang_landriault_li_2021}. 
Those studies succeed to obtain the the Laplace transform of the Gerber-Shiu function, or even the Gerber-Shiu function itself in explicit form by the inverse Laplace transform.


\subsubsection{Advanced models}

The Laplace transform of the Gerber-Shiu function has also been investigated for the Sparre-Andersen model (Section \ref{sec_Sparre-Andersen}). 
The interclaim time distributions in this context include Erlang(2) \cite{cheng_tang_2003, SUN2005}, 
a mixture of exponential distributions 
\cite{Gerber_Shiu_2005}, a mixture of an exponential distribution and an Erlang(n) distribution \cite{huang_yu_2014}, 
$K_n$ distribution \cite{Li_Garrido_2005_2}, phase-type distributions \cite{ji_zhang_2010}, and two classes of claims whose counting processes are independent Poisson and generalized Erlang(n) processes \cite{zhang_li_yang_2009}.
A generalized Erlang(n) risk process with perturbation is investigated in \cite{Li_Garrido_2005}. 
As for the claim size, one finds models with arbitrary interclaim time distribution and a Coxian claim-size distribution \cite{landriault_willmot_2007},
and phase-type interclaim times and claim-size distributions in the rational family \cite{song_meng_wu_ren_2010} with an additional multi-layer dividend strategy \cite{jiang_yang_li_2012}. 
Additional features have been introduced into the risk model, such as dependent interclaim times and claim amounts with perturbation \textcolor{black}{\cite{adekambi_2022, zhang_yang_yang_2012}}, stochastic premium income \cite{ZHAO2012}, stochastic interest force modeled by drifted Brownian model in combination with Poisson-Geometric process \cite{huang_yu_2009}, and two classes of claims modelled by compound Poisson and Erlang(2) renewal risk processes with a two-step premium rate under a threshold dividend strategy \cite{lu_xu_sun_han_2009}, \textcolor{black}{and a renewal risk model with delayed claim reporting time whose surplus earns a constant interest \cite{essiomle_2022}}.

On the risk model with claim times given by Markov-modulated Poisson processes, 
the Laplace transform is investigated with perturbation \cite{lutsai2007},
distributed claim amounts in the  rational family  \cite{lilu2008}, and delayed by-claims with random (exponentially distributed) premiums  \cite{shija_jacob_2016}. 
Finally, the Laplace transform of the Gerber-Shiu function has been derived in closed form for a Markov-dependent risk model \cite{Albrecher_Boxma_2005} with a  multi-layer dividend strategy and rational family claim-size distribution \cite{zhou_xiao_deng_2015} and for a MAP risk model with a threshold dividend strategy \cite{cheng_wang_2013}, \textcolor{black}{and for a Markov‐modulated L\'evy risk model with two‐sided jumps \cite{martin-gonzalez_murillo-salas_panti_2022}.}


\subsection{Defective renewal equations} \label{section closed - Defective renewal equation}

The integro-differential equation \eqref{Laplace 2} can be rewritten as a (defective) renewal equation.
In what follows, we review relevant results on a variety of risk models.

\subsubsection{Cram\'er-Lundberg model}

Along the line of Section \ref{section closed - Laplace transform}, we review the derivation of defective renewal equations \cite{Gerber_Shiu_1998} with a focus on the Cram\'er-Lundberg model \eqref{classical process}.
First, we multiply the integro-differential equation \eqref{Laplace 2} with $e^{-\Phi(\delta) u}$, differentiate and rearrange terms to arrive at the exponentially-tilted Gerber-Shiu function $\phi_{\Phi(\delta)}(u):=e^{-\Phi(\delta) u}\phi(u)$, that solves the integro-differential equation: 
\begin{equation}\label{DRE 1} 
	c\phi_{\Phi(\delta)}'(u)=  \lambda \mathbb{E}\left[e^{-\Phi(\delta) Y_1}\right] \phi_{\Phi(\delta)}(u)-\lambda\int_0^u\phi_{\Phi(\delta)}(x) e^{-\Phi(\delta)(u-x)}p_Y(u-x)dx-\lambda e^{-\Phi(\delta) u}\int_0^{+\infty}w(u,x)p_Y(u+x)dx,
\end{equation} 
where we have applied the definition of the solution $\Phi(\delta)$ to the Lundberg equation \eqref{Lundberg equation 1}.
We integrate the integro-differential equation \eqref{DRE 1} on the interval $[0,z]$ for $z\ge 0$ and divide by $\lambda$ to obtain 
\begin{equation}\label{DRE 2} 
	\frac{c}{\lambda}\left(\phi_{\Phi(\delta)}(z)-\phi_{\Phi(\delta)}(0)\right)=\int_0^z \phi_{\Phi(\delta)}(x) \int_{z-x}^{+\infty} e^{-\Phi(\delta) y} p_Y(y)dy dx- \int_0^z e^{-\Phi(\delta) u}\int_0^{+\infty}w(u,x)p_Y(u+x)dxdu. 
\end{equation}
In the limit $z \to + \infty$, the first terms in both sides of \eqref{DRE 2} vanish, hence resulting in 
\begin{equation}\label{DRE 3} 
	\phi_{\Phi(\delta)}(0)=\frac{\lambda}{c} \int_0^{+\infty} e^{-\Phi(\delta) u}\int_0^{+\infty}w(u,x)p_Y(u+x)dxdu. 
\end{equation}
By substituting \eqref{DRE 3} into \eqref{DRE 2} and then multiplying by $e^{\Phi(\delta) z}$ to cancel out the exponential tilting in $\phi_{\Phi(\delta)}$, we get the defective renewal equation 
\begin{equation}\label{DRE 4} 
	\phi(z)=\frac{\lambda}{c}\int_0^z\phi(x)\int_{z-x}^{+\infty}e^{\Phi(\delta)(z-x-y)}p_Y(y)dydx+\frac{\lambda}{c}\int_z^{+\infty}e^{\Phi(\delta)(z-u)}\int_0^{+\infty}w(u,x)p_Y(u+x)dxdu=:(\phi\ast g)(z)+h(z),   
\end{equation}
where $g(z):=(\lambda/c)\int_z^{+\infty}e^{-\Phi(\delta)(y-z)}p_Y(y)dy$ and $h(z):=(\lambda/c)\int_z^{+\infty}\int_0^{+\infty}e^{-\Phi(\delta)(u-z)}w(u,x)p_Y(u+x)dxdu$ for $z\ge 0$.
Its solution $\phi$ can then be computed using renewal theoretic methods.
For example, it
can be written as Neumann series 
\begin{equation}\label{DRE 5}
	\phi(u)=\sum_{k=0}^{+\infty} h \ast g^{\ast k}(u), \quad u\in (0,+\infty),
\end{equation}
where $g^{\ast k}$ denotes the $k$th order convolution for the function $g$.  
By solving the defective renewal equation \eqref{DRE 4}, the solution is found in terms of a compound geometric distribution in \cite{Lin_Willmot_1999}, as well as 
closed form expressions are derived in \cite{lin_willmot_2000moments} for several ruin quantities, such as the joint and marginal moments of the time of ruin, the surplus before the time of ruin, and the deficit at the time of ruin.

\subsubsection{L\'evy risk models}

Defective renewal equations have been derived for the Gerber-Shiu function on L\'evy risk models (Section \ref{section Levy risk model}) and their perturbations (Section \ref{section_diffusion_perturbation}) as well,
such as those involving perturbation \cite{TSAI200251}, a threshold dividend strategy \cite{Lin_Pavlova_2006}, a multi-threshold dividend strategy \cite{Lin_Sendova_2008} with additional perturbation \cite{cheung_liu_2022, yang_zhang_2009}, two-sided jumps \cite{Zhang_Yang_Li_2010}, random premium modeled by a Poisson or compound Poisson process \cite{BAO2006, LABBE2009}, along with delayed claims \cite{zhu_huang_yang_zhou_2014}, constant interest and liquid reserves \cite{cai_feng_willmot_2009}, delayed claims induced by main claims \cite{XIE2011, zou_xie_2012}, impulsive dividends \cite{LIU2015}, and stochastic discount rates modeled by Poisson or Gaussian processes \cite{wang2017}.  
The dependence structure between interclaim times and claim amounts is investigated in \cite{Boudreault2007} and also in  \cite{Zhang_Yang_2011} for the case of additional perturbation. Other examples include a constant dividend barrier and stochastic premium modeled by a compound Poisson process with exponentially distributed claim amounts \cite{zou_gao_xie_2014} and a multi-layer dividend strategy \cite{xie2017}.  
Further variants can be found, such as \eqref{EDPF moments of ruin time} with the $n$-th moment of ruin time \cite{ruan_yu_song_sun_huang_yu_2019} and \eqref{EDPF second last} incorporating the surplus level immediately after the second last claim before ruin $U_{T_{N_{\tau}-1}}$ \cite{CKCheung2010}. 
For perturbed compound Poisson risk models, a defective renewal equation of the simpler version \eqref{EDPF early version} is derived in \cite{Gerber_Landry_1998}. 
Explicit expressions for several ruin quantities are derived in \cite{ren_2005} by solving this defective renewal equation.  
An extension of the Gerber-Shiu type function \eqref{shrunk edpf} is found \cite{liu_zhang_2015} to satisfy a suitable defective renewal equation.

For a perturbed risk process driven by a subordinator, the Gerber-Shiu function satisfies a defective renewal equation solved by a Laplace transform method \cite{MORALES2007}. 
For a risk process driven by a subordinator, the renewal equation of the Gerber-Shiu function is obtained and its solution is expressed as an infinite series in  \cite{Garrido_Morales_2006}.
In addition, the case with both positive and negative jumps is studied in  \cite{zhao_yin_2010}. 
Those findings are further extended to more general L\'evy risk models in \cite{BIFFIS201092}. 
A defective renewal equation for the expression \eqref{EDPF general total cost} is obtained for a spectrally negative L\'evy risk model \cite{feng_shimizu_2013}. 

\subsubsection{Sparre-Andersen models}


For the Sparre-Andersen model (Section \ref{sec_Sparre-Andersen}), defective renewal equations have been derived when the interclaim times are  Erlang(2) \cite{cheng_tang_2003}, Erlang(n) \cite{Li_Garrido_2004}, and $K_n$ distributed \cite{Li_Garrido_2005_2}.
Defective renewal equations can be found for the compound binomial model \cite{Landriault2008} and for a general Sparre-Andersen model with both positive and negative claim values \cite{LABBE2011}.   
Renewal type equations are derived even when additional structural features are introduced, for example,  
\textcolor{black}{a delayed renewal model \cite{woo_2010}, a discrete-time renewal process with time-dependent claim sizes \cite{woo_2012},} 
a Sparre-Andersen model involving a multi-layer dividend policy \cite{YANG2008}, random claims or income modeled by a compound Poisson process \cite{CHADJICONSTANTINIDIS2009, ZHAO2012} along with perturbation and dependence between interclaim times and claim amounts \textcolor{black}{\cite{adekambi_2022, zhang_yang_yang_2012}}, a two-step premium rate under a threshold dividend strategy \cite{lu_xu_sun_han_2009}, and a stochastic discount rate modeled by a Brownian motion and a Poisson-Geometric process \cite{huang_yu_2009}. 
Defective renewal equations are derived for variants of the Gerber-Shiu function driven by a Sparre-Andersen process, such as \eqref{shrunk edpf} with general interclaim times \cite{cheung_2013}, the formulation \eqref{EDPF second last} with Coxian-distributed interclaim times and mixed Erlang-distributed claim amounts \cite{Willmot_Woo_2010} or with two-sided jumps \cite{Zhang_Yang_2010}, and the formulation \eqref{EDPF four} with dependent interclaim times and claim sizes \cite{CHEUNG2010}. 

\subsubsection{MAP risk models}

For MAP risk models (Section \ref{subsubsection markov additive processes}), defective renewal equations are obtained and solved for the ordinary Gerber-Shiu function with claim amounts of phase type that can be correlated with the interclaim times \cite{AHN2007}, for the variant \eqref{EDPF four} including the last minimum before ruin $\underline{U}_{\tau-}$ defined in \eqref{running_min} and the surplus level immediately after the second last claim before ruin $U_{T_{N_{\tau}-1}}$ \cite{cheung_landriault_2009}, for the quantity \eqref{EDPF general total cost}, which can be rewritten as the Gerber-Shiu function if the cost function is set appropriately \textcolor{black}{\cite{cheung_feng_2013, feng_shimizu_2014}}. 



\subsection{Boundary value problems
}\label{section closed - Solve PIDE}


Integro-differential equations with suitable boundary conditions can be employed for studying dividend barriers (Section \ref{Drift component}).
Here, we review a representative example \cite{Lin_Willmot_Drekic_2003} of the Gerber-Shiu function, which we denote by $\phi_b(u)$, on the Cram\'er-Lundberg model \eqref{classical process} driven by the reflected process \eqref{constant dividend surplus process} with a constant dividend barrier $b > 0$. 
First, by conditioning on the time of the first claim, we obtain the integral equation
\begin{equation}\label{BVP 1}
	\phi_b(u)=\int_0^{(b-u)/c}\lambda e^{-(\lambda+\delta)t}\gamma_b(u+ct)dt+\int_{(b-u)/c}^{+\infty} e^{-(\lambda+\delta)t}\gamma_b(b)dt, \quad u \in[0,b],
\end{equation}
where $\gamma_b(t):=\int_0^t \phi_b(t-y)p_Y(y)dy+\zeta(t)$ and $
\zeta(t):=\int_t^{+\infty} w(t,y-t)p_Y(y)dy$. 
Differentiating \eqref{BVP 1} with respect to $u$ yields
\begin{equation}\label{BVP 2}
	\phi'_b(u)=-\frac{\lambda}{c}\int_0^{u}\phi_b(u-y)p_Y(y)dy+\frac{\lambda+\delta}{c}\phi_b(u)-\frac{\lambda}{c}\zeta(u), \quad u\in (0,b).
\end{equation}
To complete this integro-differential equation properly, we impose the Neumann boundary condition $\phi'_b(b)=0$.

A variety of boundary conditions can be found in the literature embedded in the surplus process.
In the Cram\'er-Lundberg model \eqref{classical process}, relevant Gerber-Shiu functions can be represented as suitable boundary value problems in \cite{Lin_Willmot_Drekic_2003}, \cite{Albrecher2005, ma_liu_2011} and \cite{Albrecher_Hartinger_2007}, respectively, with constant, linear and multi-layer dividend barriers in accordance with the formulations given in Section \ref{Drift component}.
In particular, solutions are available explicitly for exponential/hyper-exponential claim sizes with some penalty functions, for instance, $w(x,y)=1$. 
An extension to debit interest and tax payments \eqref{eq_ming_wang} 
is investigated in \cite{Ming_Wang_Xiao_2010}, where the Gerber-Shiu function is expressed in closed form on some penalty functions and claim size distributions. 

With a constant dividend barrier, the Gerber-Shiu function is investigated on the Cram\'er-Lundberg model \cite{Lin_Pavlova_2006, Lin_Willmot_Drekic_2003} via the dividends-penalty identity $\phi_b(u)=\phi(u)-\phi'(u)V_b(u)$, 
where 
$V_b(u)$ denotes the expected present value of all dividends paid until ruin.
The dividends-penalty identity is also valid for the Cram\'er-Lundberg model with constant interest \cite{yuen_wang_li_2007}, a two-step premium rate \cite{zhang_zhou_guo_2006}, a stationary Markov risk process with only downward jumps \cite{Gerber_Lin_Yang_2006}, and a MAP risk model \cite{lu_li_2009} 
with additional perturbation \cite{cheung_landriault_2009perturbed} or a special claim amount distribution \cite{liu_xu_hu_2010}. 

A combination of credit and debit interests \eqref{eq_LI2013} 
is studied in \cite{LI2013}, where an analytic expression is derived for the Gerber-Shiu function of the simpler type (that is, $w(x,y)=w(y)$) with exponential claim amounts. 
In the Cram\'er-Lundberg model with randomized observation periods \cite{Albrecher2013}, the Gerber-Shiu function solves a certain boundary value problem and is found in closed form for claims with rational Laplace transform, again when the penalty function $w(x,y)$ is independent of $x$.
Various structural features have been investigated in this context, for example, additional perturbation and a constant dividend barrier with claim sizes of rational transform
\cite{Li_Wu_2006}, with a two-step premium rate allowing business after ruin occurs \cite{gao_2021}, perturbation and a constant interest with a periodic barrier dividend strategy \cite{yang_deng_2021}, claim amounts of phase type \cite{YTChen2007}, and a constant interest when the penalty function is independent of $x$ \cite{WANG2008}.
Other features investigated include an additional stochastic premium modeled by a compound Poisson process with multiple layers \cite{ragulina_2019}, a hybrid dividend strategy \cite{didu_2021}, a stochastic interest modeled by a drifted Brownian motion and a compound Poisson process \cite{wang_xu_yao_2007}, and stochastic volatility driven by an Ornstein-Uhlenbeck process
\cite{Chi_Jaimungal_Lin_2010}.  
\textcolor{black}{Moreover, optimal reinsurance strategies are formulated as a minimization problem of the Gerber-Shiu function, which corresponds to a unique solution of the Hamilton–Jacobi–Bellman equation with boundary conditions \cite{preischl_thonhauser_2019}. }

Finally, an algebraic operator approach is developed to solve the boundary value problem leading to an analytic expression for the Gerber-Shiu function on a Sparre-Andersen model \cite{ALBRECHER2010}, and has been extended to tackle the integro-differential equation with variable-coefficients for a Sparre-Andersen model with a surplus dependent premium rate when both claim and interclaim time distributions have rational Laplace transforms  \cite{albrecher_Constantinescu_Palmowski_Regensburger_2013}. 
Yet other boundary value problems can be found in \cite{zhanglili_2021} for an Erlang(n) risk model with two-sided jumps and a constant dividend barrier and in \cite{li_ma_2016} for a MAP risk model with a multi-layer dividend strategy and constant interest and debit rates.

\subsection{Scale functions}\label{section closed - Scale functions}

The scale function has become an indispensable tool for spectrally negative L\'evy risk models (Section \ref{section Levy risk model}).
Recall that the Laplace exponent of a spectrally negative L\'evy risk process $\{U_t:\,t\ge 0\}$ is defined by $\Psi_U(\theta):=\log\mathbb{E}[e^{\theta U_1}]$ and $\Phi(\delta)$ denotes the unique non-negative solution to the Lundberg equation \eqref{Lundberg equation}, satisfying $\Psi_U(\Phi(\delta))=\delta$.
For every $\delta\ge 0$, the $\delta$-scale function $W^{(\delta)}:\mathbb{R}\to[0,\infty)$ is defined as $W^{(\delta)}(x)=0$ if $x<0$ and otherwise an  increasing and continuous function on $[0,+\infty)$ satisfying 
\[
\int_0^{+\infty} e^{-\theta x}W^{(\delta)}(x)dx=\frac{1}{\Psi_U(\theta)-\delta},\quad \theta>\Phi(\delta).
\]
The Gerber-Shiu function can be interpreted as a generalization of the scale function.
In contrast to the scale function $W^{(\delta)}$, which is a $\delta$-harmonic function for the process $U$ (stopped at the first entrance into the negative half-line) and is zero valued on the negative half-line, the Gerber-Shiu function is an inhomogeneous version of the scale function with nonzero penalty on the negative half-line \cite{avram2021optimizing, avram2015}.
We refer the reader to \cite{kuznetsov_2013} for a comprehensive review of the scale function.

In the context of the Gerber-Shiu function, the scale function plays an important role, because the Gerber-Shiu function can be expressed in terms of the scale function for the spectrally negative L\'evy risk model \cite{BIFFIS201085}.
For instance, the scale function is rich enough to incorporate the last minimum of the surplus before ruin: 
\begin{equation} \label{GS_density}
	\phi(u)=\mathbb{E}_u\left[e^{-\delta \tau}w\left(U_{\tau-}, |U_{\tau}|, \underline{U}_{\tau-}\right) \mathbbm{1}\left(\tau<+\infty\right)\right]=\int_{(0,+\infty)^3} \mathbbm{1}\left(x\geq z\right) w(x,y,z) K_u^{(\delta)}(dx,dy,dz),
\end{equation}
where the measure in \eqref{GS_density} is defined by
\[
K_u^{(\delta)}(dx,dy,dz) :=e^{\Phi(\delta)(x-z)}\left[(\partial/\partial x)W^{(\delta)}(u-z)-\Phi(\delta)W^{(\delta)}(u-z)\right]\Pi(dy+x)dxdz.
\]
Here, the derivative $(\partial/\partial x)W^{(\delta)}$ is understood to be the right derivative (which always exists) when the left derivative does not exist.
Hence, the evaluation of the resulting Gerber-Shiu function may require numerical approximation, such as the computation of the scale function $W^{(\delta)}(x)$ and then its numerical integration \eqref{GS_density} \cite{avram2015, BIFFIS201085, library979190}. 



The scale function can be obtained by the partial fraction decomposition when the Laplace exponent $\Psi_U(\theta)$ is a rational function, such as the cases of finite activity with claim sizes of phase type \cite{BIFFIS201085, kuznetsov_2013}. 
Moreover, the scale function can be represented as an infinite series for certain risk models, such as the Cram\'er-Lundberg model perturbed by an independent Brownian motion \cite{Landriault_Willmot_2020} and $\alpha$-stable motion \cite{KOLKOVSKA2016}, and a meromorphic spectrally negative process \cite{kuznetsov_morales_2014}.
Other surplus processes on which the scale function can be expressed in closed form are 
tempered stable process and spectrally negative L\'evy processes without Gaussian components \cite{BIFFIS201085, kuznetsov_2013}. 

The Gerber-Shiu function can be written in terms of the scale function for the reflected Cram\'er-Lundberg risk model with a constant dividend barrier \cite{zhou_2005}, and the Cram\'er-Lundberg risk model perturbed by a spectrally negative $\alpha$-stable motion \cite{KOLKOVSKA2016}. 
For spectrally negative L\'evy risk processes with completely monotone L\'evy densities, the Gerber-Shiu function can be approximated by fitting hyper-exponential distributions \cite{kuznetsov_2013, yamazaki_2017}.

Special structures covered in the L\'evy risk model include Parisian ruin \cite{loeffen_Palmowski_Surya_2018}, the finite time problem \eqref{EDPF finite} in \cite{library979190}, those involving the running infimum $\underline{U}_{\tau-}$ in \cite{kuznetsov_morales_2014}, and a total cost up to default \eqref{EDPF general total cost} in \cite{feng_shimizu_2013}, where the Gerber-Shiu function is expressed in terms of the scale function.
Further extensions of the standard Gerber-Shiu function is investigated in \cite{wang2019} on spectrally negative L\'evy processes, with the ordinary ruin time \eqref{ruin_time} replaced with a drawdown time, or the moment when the surplus process declines from the peak to the subsequent trough.
In there, various quantities are expressed in terms of the scale function, such as the joint distribution of the drawdown time, the running maximum and last minimum before drawdown, and the surplus before and at drawdown.
Both Parisian ruin and drawdown time are investigated in \cite{zhang_2021} for a spectrally negative L\'evy risk model. 

Finally, a matrix version of the scale function, called the scale matrix, has been employed for MAPs in an analogous manner to the scale function for L\'evy processes \cite{feng_shimizu_2014, Ivanovs_Palmowski_2012, Kyprianou_Palmowski_2008}.
The scale matrix can be evaluated analytically in some instances \cite{ivanovs_yamazaki_2021}. 
\textcolor{black}{The (discrete-time) scale matrix for Markov additive chains is introduced in \cite{palmowski_ramsden_papaioannou_2022}.}
In addition, the scale matrix has been found useful for the computation of the Gerber-Shiu function \cite{palmowski2020phasetype} for the case with two-sided jumps using the technique known as fluid embedding.

\subsection{Recursive representations}\label{section closed -  Recursive}

On the Cram\'er-Lundberg model with a stochastic interest, the Gerber-Shiu function can be represented as an infinite series  $\phi(u)=\sum_{n\in \mathbb{N}}\phi_n(u)$, where the expected penalty function at the $n$-th claim $\phi_n(u) :=\mathbb{E}_u[e^{-\delta \tau}w(U_{\tau-},|U_{\tau}|)\mathbbm{1}(\tau=T_n)]$ can be obtained recursively \cite{Yuen_Wang_2005} as 
\[
	\phi_{n+1}(u)=\int_0^{+\infty}dx\int_0^x B(u,x)p_Y(x-y)\phi_n(y))dy,\quad n\in\mathbb{N}, 
\]
starting with $\phi_1(u):=\int_0^{+\infty}dx \int_{-\infty}^0 B(u,x)p_Y(x-y)w(x,-y)dy$, for a suitable function $B(u,x)$.
A similar approach is taken \cite{wulufang2007} for representing the Gerber-Shiu function on the Sparre-Andersen risk model with constant interest $r$ by recursion:
\begin{equation*}
	\phi_n(u)=\begin{dcases}
		\int_u^{+\infty}dx \int_{-\infty}^0 w(x,|y|) \left(\frac{r u+c}{r x+c}\right)^{\delta/r} k\left(\frac{1}{r}\log \frac{ru+c}{r x+c}\right) \frac{p_Y(x-y)}{r x+c}dy,  & \text{if } n=1,\\
		\int_u^{+\infty}dx \int_0^x \left(\frac{r u+c}{r x+c}\right)^{\delta/r} k\left(\frac{1}{r}\log \frac{ru+c}{r x+c}\right) \frac{p_Y(x-y)}{r x+c} \phi_{n-1}(y)dy, & \text{if } n\in \{2,3,\cdots\},
	\end{dcases}
\end{equation*}
for $r>0$, where $k(t)$ is the probability density function of iid interclaim times.
We refer the reader to \cite{ko_2007, wulufang2007} for further detail, also about the case $r=0$, as well as \textcolor{black}{to \cite{LEE20141, lee_willmot_2016} for dependent Sparre-Andersen models.}


For the Cram\'er-Lundberg model, a recursive relation is derived in \cite{lin_willmot_2000moments} on the $n$-th moment of the time of ruin $\mathbb{E}_u[\tau^n|\tau<+\infty]=\psi_n(u)/\psi(u)$, where $\psi_n(u)$ is given recursively by 
\[
\psi_n(u)=\frac{n}{\lambda \theta \mathbb{E}[Y]}\left[\int_0^u \psi(u-x)\psi_{n-1}(x)dx+\int_u^{+\infty} \psi_{n-1}(x)dx -\psi(u)\int_0^{+\infty}\psi_{n-1}(x)dx\right], \quad n\in\mathbb{N},
\]
starting with $\psi_0(u)=\psi(u)=\mathbb{P}_u(\tau<+\infty)$, where $\theta$ here denotes the relative security loading.
More generally, the Gerber-Shiu function involving the $n$-th moment \eqref{EDPF moments of ruin time}, say, $\phi_n(u):=\mathbb{E}_u[\tau^n e^{-\delta \tau}w(U_{\tau-},|U_{\tau}|)\mathbbm{1}(\tau<+\infty)]$, can also be represented recursively \cite{ruan_yu_song_sun_huang_yu_2019} as 
\begin{equation*}
	\phi_n(u)=\frac{n}{c}T_{\Phi(\delta)}\phi_{n-1}(u)+\frac{n}{c}\int_0^u H(u-x)T_{\Phi(\delta)}\phi_{n-1}(x)dx,\quad n\in\mathbb{N},
\end{equation*}
where $\phi_0(u)$ reduces to the standard Gerber-Shiu function \eqref{EDPF0}, and $T_sf(x):=\int_x^{+\infty}e^{-s(y-x)}f(y)dy$, which is known as the Dickson-Hipp operator. 

Finally, also for discrete models, the Gerber-Shiu function can be represented by recursion on discrete-time risk models whose surplus can be discrete as well, such as a discrete-time Sparre-Andersen model with a general interclaim time distribution \cite{wuli2009} \textcolor{black}{or with time-dependent claims \cite{woo_2012}}, a discrete-time integer-valued time-homogeneous Markov chain risk model \cite{Gerber_Shiu_Yang_2010}, a discrete-time compound binomial risk model whose claims are not identically distributed in finite time horizon \cite{bieliauskiene_siaulys_2012}, and a discrete-time compound beta-binomial model with delayed claims and randomized dividends \cite{s2019compound}.

\subsection{Distributional methods}\label{section closed -  Derivation of the Gerber-Shiu measure}

The Gerber-Shiu function can be expressed in terms of the joint distribution of random elements involved. For instance, with $f(x,y,t|u)$ the joint probability density function of the random vector $(U_{\tau-},|U_{\tau}|,\tau)$ given $U_0 = u$, the Gerber-Shiu function \eqref{EDPF0} can be written as
\[
\phi(u)=\mathbb{E}_u\left[e^{-\delta \tau}w(U_{\tau-},|U_{\tau}|)\mathbbm{1}(\tau<+\infty)\right]=\int_0^{+\infty}\int_0^{+\infty}\int_0^{+\infty} e^{-\delta t}w(x,y) f(x,y,t|u)dxdydt,
\]
so that all further analyses can focus on the joint density function.
This approach is introduced for the first time in \cite{Gerber_Shiu_1997} on the Cram\'er-Lundberg model, where its joint density function is shown to solve a certain renewal equation. 
The relevant joint density function is investigated \cite{wu_wang_wei_2003} for extensions with the time of the surplus leaving zero ultimately provided that the surplus process still continues even after it falls below zero, and is further split into two density functions depending on whether ruin is caused by the very first claim or any subsequent ones \cite{Landriault_Willmot_2009}. 
Additional structural features investigated in the literature are a constant interest \cite{Wu_Wang_Zhang_2005}, dependent interclaim times and claim amounts on the formulation \eqref{EDPF early version} \cite{Badescu_2009}, the finite time horizon \cite{dickson_2007}, and a perturbation under a multi-layer dividend strategy \cite{BRATIICHUK2012}. 
The joint distribution containing the last minimum of the surplus before ruin $\underline{U}_{\tau-}$ is investigated in \cite{Breuer2014} for a MAP risk model with phase-type jumps \textcolor{black}{and a two-sided L\'evy risk model \cite{martingonzalez_kolkovska_2022}.} 
\textcolor{black}{The discounted joint density on the last minimum of the surplus before ruin $\underline{U}_{\tau-}$ and the surplus immediately after the second last claim before ruin $U_{T_{N_{\tau}-1}}$ is discussed for a delayed renewal risk model in \cite{woo_2010}.}

As an approach based on the underlying probability measure, change of measure is applied in \cite{SCHMIDLI2010} to transform the Gerber Shiu function on Sparre-Andersen and Markov-modulated models in such a way that ruin occurs almost surely under the new probability measure. 
In this way, change of measure is useful for improving the computation of the Gerber-Shiu function via simulation (Section \ref{section simulation}). 

More distributional methods can be found in the literature for evaluating the Gerber-Shiu function. 
For instance, the Wiener-Hopf factorization is employed for deriving closed-form expressions of the formulation \eqref{EDPF early version} on a perturbed Cram\'er-Lundberg risk model with compound Poisson stochastic premium \cite{Chi_2010}. 
The finite-time Gerber-Shiu function \eqref{finite edpf 2} and some ruin times of exotic nature are investigated on the Cram\'er-Lundberg model using the (defective) density of the surplus killed on exiting \cite{LANDRIAULT2019}. 

\subsection{Asymptotic expressions}\label{subsubsection Asymptotic results}

As the availability of an analytic or semi-analytic expression for the Gerber-Shiu function is limited to a few special cases as we have described so far from place to place, its asymptotic property may often be the only useful clue in effect.
The asymptotic analysis has thus been appreciated as one of the major approaches in the context of the Gerber-Shiu function \cite{Asmussen_2010}.
In particular, on the formulation \eqref{EDPF0} with $w(x,y)=1$, it is well known that 
\begin{align}
\phi(u)\sim Ce^{-\Phi(\delta) u}, \quad \textrm{as } u\to+\infty, \label{Cramer_approx}
\end{align}
where $\Phi(\delta)$ is the unique non-negative solution to the Lundberg equation \eqref{Lundberg equation} and $C>0$, for a wide class of L\'evy risk models (Section \ref{section Levy risk model}) with light-tailed claim sizes.
The asymptotic behavior of the Laplace transform of the ruin time is found in \cite{Vsiaulys_2006} on Cram\'er-Lundberg models with 
subexponentially distributed claim sizes. 

On models with Erlang(2)-interclaim times, the asymptotic property of the moments of the deficit at ruin and of the surplus before ruin are investigated for light-tailed, subexponential, and other heavy-tailed
claim size distributions that are convolution-equivalent \cite{cheng_tang_2003}. 
In particular, the asymptotic behavior of the Gerber-Shiu function is derived in \cite{kovcetova_siaulys_2011} for an Erlang(2) risk model with classical subexponential distributed claims. 
For general Sparre-Andersen models (Section \ref{sec_Sparre-Andersen}), the asymptotics of the Gerber-Shiu function are studied on convolution-equivalent distributed claim amounts, allowing both light and heavy tails \cite{TANG2010}, and with both claim size and interclaim time distributions having rational Laplace transform, featured with the present surplus-dependent premium rate \cite{albrecher_Constantinescu_Palmowski_Regensburger_2013}. 

The asymptotic behavior of the Gerber-Shiu function on the Markov-modulated Poisson risk model is investigated in the presence of stochastic investment for light- and heavy-tailed claim size distributions \cite{ramsden_Papaioannou_2017}. 
For MAP risk models (Section \ref{subsubsection markov additive processes}), the asymptotics of the Gerber-Shiu function is studied in \cite{Albrecher_Boxma_2005}, with additional perturbation, a threshold dividend strategy and heavy-tail distributed claim amounts \cite{cheng_wang_2013}. 

We close the present section by describing a functional expansion approach \cite{pitts2007} based on the asymptotic expression \eqref{Cramer_approx}. 
Despite no analytic expression is available for the Gerber-Shiu function in most cases, 
it can be approximated in some instances in terms of the Gerber-Shiu function on a simpler risk model for which the closed-form expression is available.
There, the Gerber-Shiu function on the Cram\'er-Lundberg model \eqref{classical process} with a ``nonstandard'' claim distribution, say $H$,
is approximated in terms of the Gerber-Shiu function with exponential claim size, say $H_0$, where the constant multiple $C$ in the asymptotic expression \eqref{Cramer_approx} is to be made in common for both, that is, $\int_0^{+\infty} xe^{\Phi(\delta) x}H_0(dx) = \int_0^{+\infty} xe^{\Phi(\delta) x} H(dx) < +\infty$, by adjusting the mean of the exponential distribution $H_0$. 
Introducing the mixture distribution $H_\epsilon := \epsilon H + (1 - \epsilon) H_0$ for $\epsilon \in (0,1]$, one can construct a first-order approximation of a certain functional $\varphi$ that takes the distribution $H$ onto the corresponding Gerber-Shiu function $\phi(u)$ in the sense of 
$\epsilon^{-1}(\varphi(H_\epsilon) - \varphi(H_0)) \to \varphi'_{H_0}(H-H_0)$ as $\epsilon\to 0$,  
where the convergence holds in a Banach space which the distribution $H$ belongs to and $\varphi'_{H_0}$ denotes the Fr\'echet derivative in that space.
Based on this first-order approximation, the Gerber-Shiu function can be written as $\varphi(H) \approx \varphi(H_0) + \varphi'_{H_0}(H-H_0)$, assuming $H_\epsilon \approx H$, that is, $\epsilon$ is relatively close to $1$.
For more detail, we refer the reader to \cite{pitts2007} and references therein.

%% file: sections_ver2x/edpf_section4.tex

\section{Numerical methods}\label{section numerical methods}

The majority of the existing developments around the Gerber-Shiu function are focused on finding its analytical, semi-analytical or asymptotic representation.
Due to its complex structure, however, such analytical and semi-analytical expressions are rarely available, except for rather limited classes of penalty functions on simple risk models with elementary claim size distributions, such as exponential, Erlangs and their mixtures. 
Even if an analytical representation is available at all, it is quite often given in the form of infinite series 
or involving integrals, 
which undoubtedly require appropriate numerical techniques for the evaluation at the end in reality.



With a view towards further application of the Gerber-Shiu function in practice, we here survey the existing numerical methods, though only handful in the literature,  
on the basis of the Laplace inversion (Section \ref{section Laplace Transform}), the Fourier-cosine expansion (Section \ref{section Fourier-cosine expansion}), the numerical solution to the integro-differential equation with boundary conditions (Section \ref{section Approximate solution of integro-differential equation}), and simulation via random number generation (Section \ref{section simulation}). 
In addition, from an implementation point of view, we collectively discuss extra assumptions imposed on the surveyed numerical methods (Section \ref{subsection discussion}).

\subsection{Laplace Transform}\label{section Laplace Transform}

The inverse Laplace transform is unavailable in closed form but requires numerical approximation in many instances.
Along this line, the work \cite{library979190} derives the Laplace transform of the finite-time Gerber-Shiu function \eqref{EDPF finite} with general penalty functions in terms of the scale function (Section \ref{section closed - Scale functions}), which makes numerical inversion easier.
There, the Gaver-Stehfest algorithm is employed for numerically inverting the Laplace transform, as
\begin{equation}\label{Gaver-Stehfest formula}
\phi(u,T)\approx\frac{\log 2}{T}\sum_{k=1}^{2n} c_k \mathcal{L}_0\phi\left(u, \frac{k\log 2}{T}\right),\quad 
c_k:=(-1)^{k+n} \sum_{j=[(k+1)/2]}^{k\land n}\frac{j^n(2j)!}{(n-j)!j!(j-1)!(k-j)!(2j-k)!},
\end{equation}
where $\mathcal{L}_0\phi(u,\xi):=\int_0^{+\infty}e^{-\xi t}\phi(u,t)dt$ and $[j]$ represents the greatest integer less than or equal to $j$.
At first glance, it seems as though the quality of approximation could improve simply by increasing the truncation $n$, whereas it has been suggested on multiple occasions that the truncation $n$ is set at $[1.1m]$ for $m$ desired significant digits as well as $\sum_{k=1}^{2n}c_k=0$, as well as the system precision is set at $[2.2n]$ when the truncation $n$ is fixed, so as to achieve the expected quality of approximation.
To support the theoretical findings, numerical results are provided for various finite-time lengths in \cite{library979190}.

Another approach around the Laplace inversion is the cosine transform \cite{kuznetsov_morales_2014}, by which the finite-time Gerber-Shiu function \eqref{finite edpf 2} can be evaluated via numerical integration up to desired accuracy in the form of  
\[
\phi(u,T)=\frac{2}{\pi}e^{\delta_0 T} \int_0^{+\infty}\mathfrak{Re}\left[
\mathcal{L}_0 \phi(u,\delta_0+is) \right]\cos(sT)ds,
\]
with $\delta_0>0$.
We note, as already mentioned briefly in Section \ref{section Finite time horizon}, that the penalty function here may contain the running infimum $\underline{U}_{\tau-}$ as well.
The Gaver-Stehfest algorithm is also implemented here for comparison purposes.
It is reported that both this cosine transform and the Gaver-Stehfest algorithm are accurate and fast, yet with some pros and cons for each.
Namely, the former is robust only when the truncation is large enough and the grid is set sufficiently fine, whereas the latter can be controlled with the truncation parameter $n$ alone in \eqref{Gaver-Stehfest formula} and does not require complex numbers but higher-precision arithmetic.

\subsection{Fourier-cosine expansion}\label{section Fourier-cosine expansion}

The Gerber-Shiu function can also be approximately evaluated by truncating an infinite Fourier-cosine series.
One such method can be described as follows \cite{CHAU2015170}.
By interpreting the Gerber-Shiu function as the solution of a defective renewal equation (Section \ref{section closed - Defective renewal equation}), 
it can be represented \cite{Garrido_Morales_2006} as 
\begin{equation}\label{numerical renewal}
\phi(u)=h*\sum\limits_{k=0}^{+\infty}g^{*k}(u)=h(0)+\int_{0}^{u}V(x)dx=h(0)+\int_{0}^{a}\mathbbm{1}\left(x \leq u\right) V(x)dx,
\end{equation}
for suitable functions $h$ and $g$, depending on the surplus process, where $g^{*k}$ denotes the $k$-th order convolutions for the function $g$ and $a$ is a positive constant greater than the initial capital $u$.
Then, by applying the Fourier-cosine expansion of the integrand $V(x)=\sum\nolimits_{k=0}^{+\infty}{}^{'} A_k \cos(k \pi x/a)$, 
the representation \eqref{numerical renewal} can be rewritten as 
\begin{equation}\label{numerical renewal 2}
\phi(u)=h(0)+\int_{0}^{a}\mathbbm{1}\left(x \leq u\right) \sum\limits_{k=0}^{+\infty}{}^{'} A_k \cos\left(k \pi \frac{x}{a}\right)dx,
\end{equation}
where the notation $\sum {}^{'}$ denotes a summation with its first terms weighted by half, $A_k:=(2/a) \int_{0}^{a} V(x)\cos(k \pi x/a) dx$.
Here, the integral term can be expressed explicitly as an infinite sum using sine functions, which provides a tractable approximation by truncation of the infinite terms.
We note that a similar approach is investigated in \cite{Li_Shi_Yam_Yang_2021} to deal with the finite-time Gerber-Shiu function (Section \ref{section Finite time horizon}).

For L\'evy risk models with constant dividend barrier $b$ (Section \ref{Drift component}) and equidistant observation times $0=t_0<t_1<\cdots<t_M=T$ with stepsize $\Delta:=t_{m+1}-t_m$, the finite-time Gerber-Shiu function \eqref{finite edpf 2}, whose penalty function only contains $|U_{\tau}|$, can be represented recursively in terms of integral equations \cite{xie_zhang_2021}: for $m\in \{0,1,\cdots, M-1\}$,
\begin{align*}
\phi_m(u)&:=\mathbb{E}\left[e^{-\delta( \tau-t_m)}w\left(|U_{\tau}| \right)\mathbbm{1}\left(\tau\leq T\right)\Big|\,U_{t_m}=u, t_m<\tau\right]\\
&\approx e^{-\delta \Delta}\widehat{\phi}_{m+1}(b)\int_{-n}^{+n} \widehat{g}(u-b+y)dy+e^{-\delta \Delta}\int_{0}^b\widehat{\phi}_{m+1}(y) \widehat{g}(u-y)dy+e^{-\delta \Delta}\int_0^{+n} w(y) \widehat{g}(u+y)dy,
\end{align*}
where $n$ truncates the domain and $\widehat{g}$ denotes a suitable truncation of its Fourier-cosine expansion $\widehat{g}(x)\approx \sum_{k=0}^n {}^{'} c_k \cos(k\pi (x-a_1)/(a_2-a_1))$ on some sufficiently large truncated domain $[a_1, a_2]$ for the probability density function $g$ of stationary and independent increments $\mathcal{L}(U_{t_m-1}-U_{t_m})$. 
To justify its error analysis, an extensive collection of numerical results are presented on various distributions.

Yet another approach around the Fourier-cosine expansion is developed with the aid of the frame duality projection method \cite{wang_zhang_2019}, with which the Gerber-Shiu function for a L\'evy risk model can be approximated as  
\[
\phi(u)\approx \frac{a^{-1/2}}{\pi}\sum_{k=1}^{ab} \mathfrak{Re}\left[\int_0^{2\pi b}e^{-ix_k s}\mathcal{F}\phi_m(s)\mathcal{F}\widetilde{\varphi}(s/a)ds\right]\varphi_{a,k}(u)\mathbbm{1}\left(u\in[0,U]\right),
\]
where $a$ and $b$ are sufficiently large powers of 2, $x_k=x_1+(k-1)/b$ with negative $x_1$ according to the application, $\mathcal{F}\phi_m$ denotes the Fourier transform of $\phi_m$ (a smooth extension of the Gerber-Shiu function $\phi$), $\mathcal{F}\widetilde{\varphi}$ is the Fourier transform of the dual $\widetilde{\varphi}$ of a real-valued symmetric Riesz generator $\varphi$ with compact support, and $\{\varphi_{a,k}\}_{k\in\mathbb{Z}}$ denotes a suitable Riesz basis with scale $a$ and shift $x_k$. 
Error analysis is conducted in terms of the parameters $a$ and $b$, supported by numerical results on the Cram\'er-Lundberg model with exponential and gamma claims.
The finite-time Gerber-Shiu function for a L\'evy risk models under periodic observation is also approximated recursively via the frame duality projection \cite{xie_zhang_2022}.

\textcolor{black}{In relation to orthogonal expansions, the Laguerre polynomials starts receiving attention as an effective means for approximating (no longer only ruin probabilities, such as \cite{ALBRECHER202296, doi:10.1080/03461238.2022.2089051, doi:10.1080/03461238.2021.1885483}, but also) the Gerber-Shiu function \cite{xie_yu_zhang_cui_2022}, while the Laguerre series expansion has often been employed for obtaining the estimator of the Gerber-Shiu function.
We refer the reader to Section \ref{section inference - series expansion} for such statistical use.}

\subsection{Numerical solution of integro-differential equations}\label{section Approximate solution of integro-differential equation}

As discussed in Section \ref{section closed - Solve PIDE}, the Gerber-Shiu function can often be formulated in the form of integro-differential equation, which certainly offers various approaches of numerically solving the equation for the Gerber-Shiu function \cite{risks8010024}.
Along this line, the Chebyshev polynomial approximation has attracted a great deal of attention in dealing with Markov-modulated jump-diffusion processes \cite{DIKO2011126}, where the Laplace-Carson transform in time of the penalty-reward Gerber-Shiu function \eqref{EDPF PR early} is expanded and truncated as $\int_0^{+\infty} \xi e^{-\xi t} \phi(u,t)dt \approx \sum_{k=0}^n \xi_k \Gamma_k(u)$ with shifted Chebyshev polynomials of the first kind $\{\Gamma_k\}_{k\in\mathbb{N}_0}$ and suitable coefficients $\{\xi_k\}_{k\in\mathbb{N}_0}$ for sufficiently large truncation $n$, followed by numerical inversion of the Laplace transform. 
The Chebyshev polynomial approximation is also employed elsewhere, such as for a perturbed Markov-modulated risk model with two-sided jumps \cite{dong_zhao_2012} and Markov-dependent risk models with multi-layer dividend strategy \cite{zhou_xiao_deng_2015} (Section \ref{Drift component}).
Another approach, yet in a similar direction, is based on Stenger's sinc method \cite{zhuo_yang_chen_2018} for a phase-type risk model with stochastic investment return  and random observation periods along with encouraging numerical illustrations. 

Conceptually different approaches based on the integro-differential equation are proposed in \cite{insp, doi:10.1137/110841497}, where deterministic upper and lower bounding functions of, not an approximation of, its solution can be found numerically using semidefinite programming, or even by elementary operations alone \cite{kawai1}, for a wide variety of ruin-related quantities, supported by various numerical results.
It is worth mentioning that non-trivial boundary conditions, such as the Neumann, Robin and mixed boundary conditions are also within the scope. 

\subsection{Simulation}\label{section simulation}

Simulation via random number generation is a direct approach in evaluating the Gerber-Shiu function.
In the infinite-time situation where ruin may not occur for good, the time truncation is rather necessary than convenient for the simulation based evaluation. 
This issue can be addressed by forcing the underlying dynamics to ruin almost surely via a suitable exponential change of measure for the Sparre-Anderson, Markov-modulated, Bj\"ork-Grandell, and perturbed risk models \cite{SCHMIDLI2010}, as well as compound renewal risk models with dependence (between interclaim times and claim amounts) \cite{COSSETTE_2015}, where
the Gerber-Shiu function can be reformulated under a suitable probability measure, say, $\mathbb{P}_u^{\delta}$, as in $\phi(u)=\mathbb{E}_u[e^{-\delta \tau}w(U_{\tau-},|U_{\tau}|) \mathbbm{1}(\tau<+\infty)]=e^{-\Phi(\delta) u}\mathbb{E}^{\delta}_u[w(U_{\tau-},|U_{\tau}|) e^{\Phi(\delta) U_{\tau}}]$,
which is now free of the indicator $\mathbbm{1}(\tau<+\infty)$ inside the expectation, that is, the ruin time $\tau$ is almost surely finite under the probability measure $\mathbb{P}_u^{\delta}$.
It is also worth noting that this exponential change of measure has the potential to reduce the estimator variance in the sense of importance sampling by shortening the length of sample paths until ruin. 

Another approach through random number generation is the Wiener-Hopf Monte Carlo simulation technique \cite{ferreiro-castilla_van_schaik_2015}, where not only the ruin time but also the overshoot, undershoot and last minimum  can be sampled approximately yet jointly (such as for the formulation \eqref{EDPF four}) with the aid of stochastic exponential grid.
Error analysis is conducted with respect to the increasing exponential rate for the grid, supported by numerical results on the $\beta$-family of L\'evy processes.

\subsection{Practical aspects}
\label{subsection discussion}

A crucial step of numerical method selection is to understand what assumptions are required to verify for employing a certain numerical method. 
Some prerequisites are explicitly indicated by the method itself to be implemented, such as the closed form of the Laplace or Fourier transform,
or some knowledge on the law of the surplus process, while there may often exist extra conditions so that the numerical method can achieve desirable accuracy.
Here, we collect and briefly discuss such extra assumptions and requirements from a wider perspective.





To implement the frame duality projection \cite{wang_zhang_2019}, the Gerber-Shiu function is required to be $m$-time differentiable for some $m\in \{0,1,2\}$.
In addition, the right limit of its $k$-th derivative $\phi^{(k)}(0+)$ at the origin must exist with further integrability condition $\phi^{(k)}\in  L^1 \cap L^2(0,+\infty)$ for all $k\in \{0,\cdots,m\}$.
In fact, it might not be very realistic that all those conditions are verifiable ahead of implementation.
Furthermore, for its error analysis to fully make sense, the Gerber-Shiu function is assumed to decay faster than exponential as $\int_0^{+\infty}e^{ru}\phi(u)du<+\infty$ for some positive $r$.

In the Fourier-cosine expansion method \cite{CHAU2015170}, the non-trivial function $V(x)$ in the representation \eqref{numerical renewal}, which involves convolutions and infinite series, is assumed to be in $L^1 \cap L^2(0,+\infty)$.
In addition, for its error analysis to hold fully true, the real part of the Fourier transform of the integrand $V(x)$ is assumed to have an algebraic index of convergence of some positive $\beta$, that is, there exists a positive $\beta$ satisfying $\limsup_{k\to +\infty}|\mathfrak{Re}(\int_0^{+\infty}e^{ik\pi x}V(x)dx)|k^{\beta}<+\infty,$ with which error bounds are written in terms of finite truncation of each of the infinite Fourier-cosine expansion and the semi-infinite region of the Fourier transform.
The finite-time problem is investigated \cite{Li_Shi_Yam_Yang_2021} on a similar set of conditions.

To apply numerical methods based on the integro-differential equation, its solution is assumed to satisfy suitable smoothness conditions on its own, such as twice differentiability \cite{dong_zhao_2012, zhuo_yang_chen_2018}, continuous second derivatives \cite{DIKO2011126} and continuity at every dividend barrier \cite{zhou_xiao_deng_2015}. 
As usual in this context, moreover for the solution to be smooth, suitable regularity and integrability conditions are imposed on the problem elements, such as the claim size distribution, the reward, volatility and penalty functions \cite{DIKO2011126}, and 
the premium rate \cite{zhou_xiao_deng_2015}. 

As for the simulation based methods \cite{COSSETTE_2015, ferreiro-castilla_van_schaik_2015}, it is indeed a major advantage that no extra regularity conditions are required,.
In particular, unlike the other numerical methods, the evaluation of the expectation does not even require, for instance, continuity in the initial surplus.
We note that the convergence of simulation based methods 
is generally guaranteed, provided that sample paths of the surplus process can be simulated exactly, or almost exactly even in the presence of jumps of infinite activity \cite{10.1214/20-PS359} by Monte Carlo methods.
Moreover, the method via change of measure \cite{COSSETTE_2015} may be thought of as a variance reduction technique of importance sampling, which may often accelerate the weak convergence in terms of estimator variance. 
For the Wiener-Hopf Monte Carlo simulation technique  \cite{ferreiro-castilla_van_schaik_2015}, two sources of error and convergence need to be addressed, not only from the strong law of large numbers, but also from a discretization of the time interval.
The convergence on each source has been proved, whereas the latter error seems difficult to track in general for the Gerber-Shiu function.  

Having discussed and contrasted a variety of advantages and disadvantages all the way through the present section, we stress that there is ultimately no such thing as a perfect numerical method that works best under least conditions in every single problem setting.
For instance, on the one hand, infinite-time problems, or equivalently elliptic integro-differential equations, can be addressed often very efficiently by deterministic numerical methods (Sections \ref{section Laplace Transform}, \ref{section Fourier-cosine expansion} and \ref{section Approximate solution of integro-differential equation}), whereas those stand on a variety of strict conditions. 
On the other hand, simulation based methods (Section \ref{section simulation}) are at least implementable under few conditions in virtually all problem settings, whereas those are not a good choice for infinite-time problems because one needs to keep generating every sample path until its ruin.
Indeed, one might be better off adopting methods under restrictive conditions, no matter how extremely restrictive, as long as all such requirements can be verified.
As such, the numerical method needs to be very carefully selected on the basis of adequate understanding of the specific problem structure and its  conditions, also taking implementation perspectives into account, such as a priori estimated cost for coding and available computing budget.

%% file: sections_ver2x/edpf_section5.tex
\section{Statistical inference}\label{section inference}

The analytical, semi-analytical, asymptotic and numerical methods of Sections \ref{section analytical and semi-analytical approaches} and \ref{section numerical methods} address the forward problem where a risk model has been fixed with all its parameters specified in advance.
In this section, we turn to the inverse problem, that is, statistical inference for those model parameters on the basis of observed data. 
For instance, in the classical Cram\'er-Lundberg model, if the claim size distribution is known in closed form with finite dimensional parameters, then the maximum likelihood estimation and method of moments can be applied based on the data of claim numbers and sizes. 
In the literature, however, there seem to exist only a handful of studies addressing statistical inference on advanced models, such as Sparre-Andersen and Markov additive models (Sections \ref{sec_Sparre-Andersen} and \ref{subsubsection markov additive processes}), but all focused on ruin probabilities.
When it comes to the Gerber-Shiu function in general, to the best of our knowledge, all existing studies seem to base on the L\'evy risk model (Section \ref{section Levy risk model}), to which we devote the present section.
That is, we suppose throughout that the insurance surplus is modeled as  
\begin{equation}
U_t = u + c t  + \sigma W_t - L_t, \label{si:surplus}
\end{equation}
where $u\ge 0$, $c\in (0,+\infty)$, $\{W_t:\,t\ge 0\}$ is a Wiener process, and $\{L_t:\,t\ge 0\}$ is a subordinator with L\'evy measure $\Pi$
satisfying the so-called net profit condition $\mathbb{E}[L_1] = \int_{0+}^{+\infty} z\Pi(dz) < c$. 
We note that the net profit condition implies that the L\'evy process $\{L_t:\,t\ge 0\}$ is of bounded variation, which is a convenient property in the context of the Gerber-Shiu function and its inference for technical reasons \cite{feng_shimizu_2013, Shimizu201784}. 

\subsection{Sampling scheme}

From a statistical point of view, it is imperative to define the observation scheme for asymptotic analysis, with the chosen model \eqref{si:surplus} taken into account.

If the surplus model \eqref{si:surplus} contains the non-degenerate diffusion component of infinite variation (that is, $\sigma>0$), it is unnatural to assume that its trajectory is fully observable on any interval. 
Hence, one often assumes that the surplus $\{U_t:\,t\ge 0\}$ is only observed at equidistant discrete time points $t_k = k h_n$ for $k\in \{0,1,\cdots,n\}$ with step size $h_n>0$, that is, $(n+1)$-observations $(U_{t_0}, U_{t_1}, \cdots, U_{t_n})=:U^n$ are available. 

In the case of finite activity ($\lambda <+\infty$), every claim size $\{Y_k\}_{k\in \mathbb{N}}$ is observable.
We remark that each $Y_k$ may not be interpreted exactly as a claim size, particularly in the presence of the diffusion component of infinite variation \cite{Shimizu2009}.
If, moreover, the diffusion component is degenerate ($\sigma=0$), then the discrete time observations $U^n$ are computable based on the claims data and the premium rate $c>0$, both of which are available in the context of insurance.
If, however, the jump component is of infinite activity ($\lambda =+\infty)$, it is impossible to observe and store all jumps, where one usually assumes the availability of ``large'' jumps alone (for instance, \cite{Shimizu2011}), say, $J^{\epsilon}:=\{\Delta L_t:\, \Delta L_t > \epsilon \mbox{ for }t\in [0,t_n]\}$ for a suitable threshold $\epsilon>0$, where we have denoted $\Delta L_t:=L_t-L_{t-}$.
In summary, the available data of the trajectory can be formulated as  
\begin{align}
	U^n \cup J^{\epsilon},\label{si:data}
\end{align}
for some $n\in \mathbb{N}$ and $\epsilon>0$, with an exception of $\epsilon=0$ when the jump component is of finite activity ($\lambda<+\infty$).

With a view towards asymptotic analysis, on the one hand, one often imposes the so-called high-frequency condition $h_n\to 0$ as $n\to +\infty$ to describe the case where, for instance, daily (or even more frequent intraday) book data are available for the past years.
Indeed, the advent of technology has increased the means of such high-frequency sampling.
On the other hand, if the frequency is much lower (for instance, only \textcolor{black}{weekly} or monthly data are available), then it is more sensible to impose the low-frequency condition (for instance, \cite{WANG2022IME, Zhang_Yang_2014}) with a fixed step size $h_n\equiv h(>0)$, no matter how many observations are available.
From a technical point of view, it is often impossible to construct a (statistically) efficient estimator for unknown parameters in low-frequency scenarios.
Hence, along realistic trends towards more advanced data collection technology, high-frequency sampling tends to attract increasingly more attention than low-frequency counterpart in the context of statistical inference.
We thus devote this section to high-frequency scenarios.

If the jump component is of infinite activity ($\lambda=+\infty$), it is often the case to take the limit $\epsilon\to 0$ in the available data \eqref{si:data} along increasing observations $n\to +\infty$, representing that smaller jumps are observable as well.   
In every situation, one usually lets the observation time window expand ($t_n\to +\infty$) so as to better identify the model parameter in the jump component. 

\subsection{Estimating L\'evy characteristics}\label{subsection estimating levy characteristics}

When estimating the Gerber-Shiu function, we preliminarily need to estimate the characteristics of the underlying L\'evy process and its relevant quantities.
Hereafter, we suppose that data set is available in the form \eqref{si:data} with the conditions $h_n\to 0$, $\epsilon_n \to 0$ and $t_n\to +\infty$, as $n\to +\infty$.

The parameters of interest are (1) the diffusion coefficient $\sigma$ and (2) functionals of the L\'evy measure $\Pi(V):=\int_{0+}^{+\infty} V(z)\Pi(dz)$ for a $\Pi$-integrable function $V:(0,\infty)\to \mathbb{R}$. 
As for (1), the following estimator is convenient \cite[Theorem 2.6 (c) and Remark 5]{jacod2007}: for given $t>0$, 
\begin{equation}
\widehat{\sigma}^2_n:= \frac{1}{t}\left[\sum_{k=1}^{[t/h_n]} |\Delta_k U^n|  - \sum_{\Delta L_t \in J^{\epsilon_n}} |\Delta L_t|^2\right],   \label{est:sigma}
\end{equation}
where $\Delta_k U:= U_{t_k} - U_{t_{k-1}}$, in the sense of $\sqrt{t_n}(\widehat{\sigma}^2_n - \sigma^2) \stackrel{\mathbb{P}}{\to} 0$, as $n\to +\infty$, provided that $nh_n^2\to 0$ and $\sqrt{t_n} \int_0^{\epsilon_n} z^2\Pi(dz) \to 0$. 
As for (2), with the functional
\begin{equation}
\widehat{\Pi}_n(V):= \frac{1}{t_n}  \sum_{\Delta L_t \in J^{\epsilon_n}} V(\Delta L_t),  \label{est:pi-v}
\end{equation}
it is known \cite{Shimizu2011} that 
\begin{equation}
\sqrt{t_n}(\widehat{\Pi}_n(V) - \Pi(V)) \stackrel{d}{\to} \mathcal{N}(0, \Pi(V^2)),
\end{equation}
as $n\to +\infty$, if $\sqrt{t_n} \int_{0+}^{\epsilon_n} V^2(z)\Pi(dz) \to 0$ in addition to suitable moment conditions of the L\'evy measure $\Pi$.
We note that the condition $\sqrt{t_n} \int_{0+}^{\epsilon_n} V^2(z)\Pi(dz) \to 0$ requires that the threshold $\epsilon_n$ decrease in $n$, due to the expanding observation window $t_n\to +\infty$. 
We note that high-frequency sampling may cause some technical issues, such as limiting singularity of the Fisher information matrix, in estimating characterizing parameters of the relevant L\'evy measure $\Pi$ \cite{10.1007/978-3-642-33134-3_87}.

With the aid of the estimators \eqref{est:sigma} and \eqref{est:pi-v}, one can estimate the Lundberg exponent $\rho$ as the absolute value of the smallest non-positive solution to the Lundberg equation \eqref{Lundberg equation} in the case of finite activity ($\lambda < +\infty$), or as the largest non-negative solution to the generalized Lundberg equation: 
\[
\ell(x; \sigma,\Pi):= cx + \frac{\sigma^2}{2} x^2 + \int_{0+}^{+\infty} (e^{-x z} - 1)\Pi(dz) - \delta = 0,
\]
in the case of infinite activity ($\lambda = +\infty$).
Note that the premium rate $c$ is a given constant in the context of insurance. 
One can then construct an $M$-estimator for the Lundberg exponent $\rho$: 
\begin{equation}
\widehat{\rho}_n := \argmin_{x \in [0,+\infty)} \ell^2\left(x; \widehat{\sigma}_n, \widehat{\Pi}_n(e^{-x\cdot} -1)\right), \label{est:rho}
\end{equation}
which is asymptotically normal around the true exponent $\rho$, such that 
\[
\sqrt{t_n} (\widehat{\rho}_n - \rho) \stackrel{d}{\to} \mathcal{N}\left(0,
\frac{\int_{0+}^{+\infty} (1 - e^{-\rho z})\Pi(dz)}{c + \sigma\rho + \int_{0+}^{+\infty} ze^{-\rho z}\Pi(dz)}\right), 
\]
as $n\to +\infty$.

\subsection{Laplace inversion}\label{section inference - Laplace inversion}

As was discussed in Section \ref{section closed - Laplace transform}, it is straightforward to perform the Laplace transform of the Gerber-Shiu function as it satisfies the defective renewal equation \eqref{DRE 1}.
In addition, when the jump component is of infinite activity ($\lambda=+\infty$), explicit expressions are available for the Laplace transform \cite{BIFFIS201092}. 
To the best of our knowledge, the Laplace inversion is indeed the first approach that achieved consistent estimators in the history of statistical inference for Gerber-Shiu functions \cite{Shimizu2011, Shimizu2012}.

In the sequel, we 
use the notation $\Pi(x;w):= \int_x^\infty w(x,z-x)\Pi(dz)$,
provided that the integral is finite valued.
By applying the expression \cite[Section 3.3]{Shimizu2011}:
\begin{align}
	{\cal L}\phi(s) = \frac{{\cal L}\Pi(\rho;w) - {\cal L}\Pi(s;w) + w(0,0)\sigma^2(s-\rho)/2}{(s-\rho)(\sigma^2(s + \rho)/2 + c) + \Psi_L(-s)- \Psi_L(-\rho)},
	\label{si:Laplace-gs}
\end{align}
one may estimate the Gerber-Shiu function $\phi$ by (numerically) inverting the following estimator for the Laplace transform \eqref{si:Laplace-gs}: 
\[
\widehat{{\cal L}\phi}_n(s) = \frac{\widehat{{\cal L}\Pi}_n(\widehat{\rho}_n;w) - \widehat{{\cal L}\Pi}_n(s;w) + w(0,0)\widehat{\sigma}^2_n(s-\widehat{\rho}_n)/2}{(s-\widehat{\rho}_n)(\widehat{\sigma}^2_n(s + \widehat{\rho}_n)/2 + c) +\widehat{\Psi}_n(-s)- \widehat{\Psi}_n(-\widehat{\rho}_n)}, 
\]
where we have employed \eqref{est:sigma}, \eqref{est:pi-v} and \eqref{est:rho}, and $\widehat{{\cal L}\Pi}_n(x;w):= {\cal L}\widehat{\Pi}_n(w(x, x-\cdot)\mathbbm{1}_{[x,+\infty)}(\cdot))$ and $\widehat{\Psi}_n(s):= \widehat{\Pi}_n(e^{s\cdot}-1)$. 
The Laplace inversion ${\cal L}^{-1}$ does not have a ``continuity", that is, even if the estimator $\widehat{{\cal L}\phi}$ satisfies consistency $\widehat{{\cal L}\phi}_n \stackrel{\mathbb{P}}{\to} {\cal L}\phi$, the Laplace inversion might not be consistent ${\cal L}^{-1}\widehat{{\cal L}\phi}_n\stackrel{\mathbb{P}}{\not\to} \phi$.
One may then employ the regularized Laplace inversion \cite{Chau1994}, defined by ${\cal L}_m^{-1}g(t):= \frac{1}{\pi^2}\int_{0+}^{+\infty} \Gamma_m(y)y^{-1/2} {\cal L}g(ty)\,dy$ for a square integrable function $g$, where $\Gamma_m(y):= \int_0^{a_m} \cosh(\pi x)\cos(x\log(y))\,dx$ and $a_m:=\pi^{-1}\cosh^{-1}(m\pi)$. 

Now, we give an estimator for the Gerber-Shiu function $\phi$: for every $\theta> 0$ and a sequence $\{m_n\}_{n\in \mathbb{N}}$ satisfying $\lim_{n\to +\infty}m_n=+\infty$, define
\begin{align}
	\widehat{\phi}^\theta_n(u) := e^{\theta u}{\cal L}_{m_n}^{-1} (\widehat{{\cal L}\phi}_n)_\theta(u)  
	=  \frac{e^{\theta u}}{\pi^2} \int_{0+}^{+\infty} {\cal L}\psi_n(us)\widehat{{\cal L} \phi}_n(s + \theta)\,ds, 
	\label{est:gs-laplace}
\end{align}
where
\[
\psi_n(x):= \frac{\pi\sinh(\pi r_n) \cos(r_n\log(x)) + \log(x)\cosh(\pi r_n)\sin(r_n\log(x))}{\sqrt{x}(\pi^2 + (\log (x))^2)},\quad r_n:= \frac{1}{\pi}|\cosh^{-1}(\pi m_n)|. 
\]
Here, we have employed the $\theta$-shift ${\cal L}\phi_\theta(\cdot):= {\cal L}\phi(\cdot+ \theta)$ because the regularized Laplace inversion ${\cal L}_m^{-1}$ may not be applied to the Laplace transform ${\cal L}\phi$ if the Laplace transform ${\cal L}\phi$ in \eqref{si:Laplace-gs} is not in $L^2(0,+\infty)$.

The rate of convergence of the estimator \eqref{est:gs-laplace} is investigated in terms of the integrated squared error \cite{Mna2008} with the norm $\| f \|_K:= (\int_0^K |f(x)|^2\,dx)^{1/2}$ for $K>0$.
Under suitable regularity conditions, take a constant $\theta$ and a sequence $\{m_n\}_{n\in\mathbb{N}}$ such that $\theta > \delta (c - \int_{0+}^{+\infty} z\Pi(dz))^{-1}$ and $m_n = {\cal O}(\sqrt{t_n\log(t_n)})\uparrow +\infty$ in \eqref{est:gs-laplace}.
Then, if $nh_n^2\to 0$ as $n\to +\infty$, then it holds that for every $\theta>0$ and $K>0$,
\begin{align}
	\|\widehat{\phi}^\theta_n - \phi\|_K = {\cal O}_{\mathbb{P}}(|\log(t_n)|^{-1/2}), \label{est:consistency-laplace}
\end{align}
where $\mathcal{O}_{\mathbb{P}}$ denotes the usual stochastic boundedness.
We note that the logarithmic rate of convergence is way too slow, relative to the standard rate ${\cal O}_p(\sqrt{t_n})$ in the context of statistical inference.
It is known \cite{Shimizu2012} that this slow rate of convergence cannot be improved even when the L\'evy measure is finite ($\lambda<+ \infty)$.
If one wishes to pursue faster rates, it is natural to adopt the Fourier inversion, which is a well-posed transform, unlike the Laplace inversion.

\subsection{Fourier inversion}\label{section inference - Fourier inversion}

In this section, we summarize statistical methods based on the Fourier inversion on the surplus model \eqref{si:surplus} without diffusion component ($\sigma = 0$) and with an absolutely continuous L\'evy measure $\Pi(dz) = \nu(z)\,dz$.
Here, we are concerned with statistical inference from the data set $U^n$ without the jump one $J^{\epsilon_n}$, following \cite{Shimizu201784}.
Note that $\Delta_kL= L_{t_k}- L_{t_{k-1}} = c h_n - \Delta_k U^n$ for all $k\in \{1,\cdots,n\}$ and is thus available only from $U^n$. 

A formula of for the Gerber-Shiu function $\phi$ via Fourier inversion is given by \cite[Theorem 2.1]{Shimizu201784}: 
\begin{align}
	\phi(u) = \frac{1}{2\pi}\int_\mathbb{R} e^{-isu}\frac{N(s)}{c - D(s)}\,ds, \label{gs:fourier-inv}
\end{align}
where $D(s):= (\Psi_L(is) - \Psi_L(-\rho))/(\rho + is)$, $N(s):= \int_{0+}^{+\infty} a(s;z,\rho)\nu(z)dz$ and $a(s;z,\rho):= \int_{0+}^{+\infty} e^{-(\rho + is)y } (\int_y^z  e^{isx}w(x,z-x)dx)dy$ with the Lundberg exponent $\rho$.
On the basis of this formula, an estimator for the Gerber-Shiu function $\phi$ can be constructed as follows: 
\begin{equation}
\widehat{\phi}_{m,n}(u):= \frac{1}{2\pi} \int_{-m\pi}^{m\pi} e^{-isu} \frac{\widehat{N}_n(s)}{c - \widehat{D}_n(s)}ds\mathbbm{1}(A_n), \label{est:gs-fourier}
\end{equation}
where $m>0$, $\widehat{D}_n(s):= (\widehat{\Psi}_n(is) - \widehat{\Psi}_n(-\widehat{\rho}_n))/(\widehat{\rho}_n + is)$, $\widehat{\Psi}_n(s)$ is as defined in Section \ref{section inference - Laplace inversion}, 
$\widehat{N}_n(s):= (nt_n)^{-1} \sum_{k=1}^n a(s;\Delta_kL, \widehat{\rho}_n)$, and $A_n:=\{ \omega\in \Omega:\, c-(t_n)^{-1}\sum_{k=1}^n \Delta_kL \ge K(\log(t_n))^{-1/2}\}$ for a constant $K>0$.
The parameter $m$ is employed for truncation since the integrand $\widehat{N}_n(s)/(c - \widehat{D}_n(s))$ is not integrable at infinity. 
This truncation parameter $m$ is then further parameterized to be an increasing sequence as $m_n\to +\infty$ for deriving consistency of the estimator \eqref{est:gs-fourier}. 
The indicate $\mathbbm{1}(A_n)$ is essential since the estimator \eqref{est:gs-fourier} might not be well defined sometimes without it.
In fact, the net profit condition $\int_{0+}^{+\infty}z\nu(z)dz<c$ plays a crucial role here for ensuring $\mathbb{P}(A_n)\to 1$ as $n\to +\infty$ for all $K>0$.


Consistency of the estimator $\widehat{\phi}_{m_n,n}$ is then described in terms of the mean integrated squared error: 
\[
\mathbb{E}\left[ \|\widehat{\phi}_{m_n,n}-\phi\|^2\right] = {\cal O}(|\log(t_n)|/t_n) + {\cal O}(h_n^\alpha |\log(t_n)|),\quad n\to +\infty,
\]
provided that $m_n\sim \sqrt{t_n/|\log(t_n)|}$, $nh_n^3 = {\cal O}(1)$, and there exists $\alpha\in (0,2)$ such that $h_n^{2-\alpha} \nu(h_n)\to 0$.
Note that 
$\alpha\in (0,2-\beta)$ with the Blumenthal-Getoor index $\beta:= \inf\{p>0:\, \int_{0+}^1 |x|^p\nu(z)dz <+\infty \} \in (0,2)$.
As a result, for every $\gamma>0$ satisfying $nh_n^{1+\gamma}\to +\infty$, it holds that  
\begin{equation}\label{convergence phi mn n}
\sqrt{nh_n^{1+ \gamma}}(\widehat{\phi}_{m_n,n}(u) - \phi(u)) \stackrel{\mathbb{P}}{\to} 0,
\end{equation}
for all $u\in [0,+\infty)$, provided that $nh_n^{1+ \alpha}\to 0$.
Indeed, this rate is a lot faster than the logarithmic rate \eqref{est:consistency-laplace} based on the Laplace transform.
We refer the reader to \cite{Shimizu201784} for relevant numerical comparisons between the two estimators \eqref{est:gs-laplace} and \eqref{est:gs-fourier}.
We close this section by noting that the exact rate (as opposed to the dominating rate \eqref{convergence phi mn n}) and limiting distribution are still unknown. 

\subsection{Series expansions}\label{section inference - series expansion}

Estimation of the Gerber-Shiu function can also be tackled via its series representation, which can often provide a means for revealing the exact rate of convergence and the limiting distribution, unlike the Laplace and Fourier inversions (Sections \ref{section inference - Laplace inversion} and \ref{section inference - Fourier inversion}).

The Fourier-cosine series expansion is first employed \cite{Yang_Su_Zhang_2019} in estimating the discounted density of the deficit at ruin $f_{\delta}(u,x)=\int_0^{\infty}e^{-\delta t}f(u,x,t)dt$ on the Cram\'er-Lundberg model (Section \ref{section Cl model}). 
The estimator of the Gerber-Shiu function on a risk model with compound Poisson random income is represented by Fourier-cosine series expansion \cite{wang_yu_huang_2019, wang_yu_huang_yu_fan_2019} (without constant premium in the latter), and on a L\'evy risk model \cite{su_wang2021, WANG2022IME} (Section \ref{section Levy risk model}).
We note that the work \cite{WANG2022IME} is focused on low-frequency scenarios ($h_n\equiv h(>0)$), under which theoretical results, such as the rate of convergence, differ from ones under high-frequency scenarios that we present in what follows.
The Fourier-sinc series expansion, equivalent to Fourier-cosine, is also investigated in \cite{Zhang_2017}. 

Now, in light of the expansion \eqref{numerical renewal 2}, the Gerber-Shiu function $\phi(u)$ for $u\le a$ can be written as 
\[
\phi(u) = \sum_{k=0}^{+\infty}{}^{'} \frac{2}{a}\mathfrak{Re}\left(\int_0^a e^{\frac{ik\pi}{a}x} \phi(x)\,dx\right) \cos\left( k\pi \frac{u}{a}\right)
\approx  \sum_{k=0}^{K-1}{}^{'} \frac{2}{a}\mathfrak{Re}\left( \frac{N(k\pi/a)}{c-D(k\pi/a)}\right) \cos\left( k\pi \frac{u}{a}\right)=:\phi^{K,a}(u), \quad u\le a, 
\]
where $N$ and $D$ are as appeared in \eqref{gs:fourier-inv}.
Therefore, with the aid of the estimators in \eqref{est:gs-fourier}, one can construct a truncated estimator for the Gerber-Shiu function $\phi$: 
\[
\widehat{\phi}^{K,a}_n(u):= \sum_{k=0}^{K-1}{}^{'} \frac{2}{a}\mathfrak{Re}\left( \frac{\widehat{N}_n(k\pi/a)}{c-\widehat{D}_n(k\pi/a)}\right) \cos\left( k\pi \frac{u}{a}\right), \quad u\le a. 
\]
Its $L^2(0,+\infty)$-error can be decomposed into the statistical and truncation terms, respectively, as: 
\[
\| \widehat{\phi}^{K,a}_n - \phi \| \le \| \widehat{\phi}^{K,a}_n - \phi^{K,a} \| + \| \phi^{K,a} - \phi \| =: E_{\text{s}} + E_{\text{n}}, 
\]
which behave like
\[
E_{\text{s}} = {\cal O}\left(\frac{K}{a}h_n^{2\alpha} + \left(\frac{K}{a}\right)^3h_n^2\right) 
+ {\cal O}\left(\frac{K}{a}\left(\frac{|\log(K/a)| +1}{t_n}+ h_n^2\right)\right), \quad
E_{\text{n}} ={\cal O}\left( \frac{a}{K}+ \frac{K}{a}\int_a^{+\infty} |\phi(x)|^2dx\right),
\]
as $K\to +\infty$, where $\alpha$ is a constant taking values in $(0,2)$ as defined in Section \ref{section inference - Fourier inversion}. 

The Laguerre series expansion is another valid candidate in estimating the Gerber-Shiu function with potential for a fast convergence rate along with the availability of the limiting law. 
To describe the approach, let $L_n$ be the Laguerre polynomial of order $n$, that is, $L_n(x):= (e^x/n!)(d^n/dx^n)(e^{-x}x^n)= (1/n!)(d/dx-1)^nx^n$, and define the sequence $\{\varphi_k\}_{k\in \mathbb{N}_0}$ by $\varphi_k(x):= \sqrt{2}L_k(2x)e^{-x}$ for $x\in [0,+\infty)$, which forms a complete orthonormal $L^2([0,+\infty))$-basis. 
Hence, every $f\in L^2([0,+\infty))$ can be expanded as $f(x) = \sum_{k\in\mathbb{N}_0} a_{f,k} \varphi_k(x)$, with $ a_{f,k}:= \langle f,\varphi_k\rangle = \int_0^{+\infty} f(x)\varphi_k(x)dx.$
A possible advantage of employing the Laguerre series expansion is that the relevant estimator can be shown to be asymptotically normal, which is first discussed in \cite{sz19} in the context of ruin probabilities under L\'evy risk models.
The Laguerre series expansion has found its application in obtaining the estimator of the Gerber-Shiu function under the Cram\'er-Lundberg model \cite{zhang_su_2018} with its asymptotic normality \cite{su_yu_2020}. 
We note that the Laguerre series expansion can be applied to more advanced surplus processes, such as a Cram\'er-Lundberg model with random income \cite{su_shi_wang_2020}, Wiener-Poisson models \cite{su_yong_zhang_2019}, pure-jump L\'evy risk models \cite{zhang_su_2019}, and spectrally negative L\'evy risk process \cite{huang_yu_pan_cui_2019}.
In light of series expansions \eqref{DRE 4}, one can approximate $\phi$, $g$ and $h$ by Laguerre series expansion, as for $m\in \mathbb{N}$, 
\[
\phi_m(x):= \sum_{k=0}^{m-1} a^{\phi}_k \varphi_k(x),\quad 
g_m(x):= \sum_{k=0}^{m-1} a^g_k \varphi_k(x),\quad 
h_m(x):= \sum_{k=0}^{m-1} a^h_k \varphi_k(x), 
\]
where, in the context of the L\'evy risk model, 
\begin{align*}
	a^g_k &:= \int_0^{+\infty}\frac{1}{c}\left[\int_0^z e^{-\rho(z-x)} \varphi_k(x)\,dx\right]\Pi(dz) =: \int_{0+}^{+\infty} Q_k(z,\rho)\Pi(dz); \\
	a^h_k &:= \int_0^{+\infty} \frac{1}{c}\left[\int_0^z\int_u^z e^{-\rho(x-u)} w(x,z-x)\varphi_k(u)\,dxdu\right]dz=: \int_{0+}^{+\infty} R_k(z,\rho)\Pi(dz),
\end{align*}
with $a^\phi_k:= \sum_{j=0}^{k-1} (a^g_{k-j} - a^g_{k-j-1}) a^{\phi}_j/\sqrt{2} + a^h_k$, and $a^\phi_0 = a^g_0 a^\phi_0/\sqrt{2} + a^h_0$.
With the notation of Section \ref{subsection estimating levy characteristics}, their consistent estimators can be constructed as  
\[
\widehat{a}^g_{k,n}:= \frac{1}{nh_n} \sum_{j=1}^n Q_k(\Delta_j L,\widehat{\rho}_n),\quad 
\widehat{a}^h_{k,n}:= \frac{1}{nh_n} \sum_{j=1}^n R_k(\Delta_j L,\widehat{\rho}_n),\quad 
\widehat{a}^{\phi}_{k,n}:= \sum_{j=0}^{k-1} \frac{1}{\sqrt{2}}(\widehat{a}^g_{k-j,n} - \widehat{a}^g_{k-j-1,n}) \widehat{a}^{\phi}_{j,n} + \widehat{a}^h_{k,n},
\]
which yields a truncated estimator for the Gerber-Shiu function $\phi$ in the form of $\widehat{\phi}_{m,n}(u):= \sum_{k=0}^{m-1} \widehat{a}^{\phi}_{k,n} \varphi_k(u)$ for $u\in [0,+\infty)$. 
As before, its error can be decomposed into two components, that is to say, the statistical error $E_{\text{s}}$ and the numerical error $E_{\text{n}}$, as follows 
\[
\| \widehat{\phi}_{m,n} - \phi\| \le \| \widehat{\phi}_{m,n} - \phi_m\| + \| \phi_m - \phi\| :=E_{\text{s}} + E_{\text{n}}=\left[{\cal O}\left(m^2\left(\frac{1}{nh_n} + h_n^2\right)\right)+o(m^2h_n^{2\alpha})\right]+  {\cal O}(m^{-r}),
\quad m,n\to +\infty,
\]
where $r:= \sup\{q>0: \sum_{k\in \mathbb{N}} k^q (a^{\phi}_k)^2 < +\infty \}$ and $\alpha$ is the same index in $(0,2)$ as before. 
By suppressing the second term of the statistical error $E_{\text{s}}$ and minimizing the remaining two terms with respect to $m$, one obtains the optimal order ${\cal O}( ((nh_n)^{-1} + h_n^2)^{-1/(r+2)})$. 
Moreover, for each degree of the expansion $m$, it has been shown that there exists a positive function $\Sigma_m:(0,+\infty)\to (0,+\infty)$ such that  
\begin{align*}
	\sqrt{t_n}(\widehat{\phi}_{m,n}(u) - \phi_m(u))  \stackrel{d}{\to} \mathcal{N}(0,\Sigma_m(u)), \quad n\to +\infty,
\end{align*}
for all $u\in [0,+\infty)$.
As for the expression of the limiting variance $\Sigma_m$, we refer the reader to \cite{su_yu_2020} for the case where $\{L_t:\,t\ge 0\}$ is a compound Poisson process, and to \cite{sz19} for the ruin probability (that is, $w\equiv 1$ and $\delta=1$).
We note that the other models, as well as the asymptotic results as $m\to +\infty$, are still open problems. 

Yet another estimation method can be found in \cite{dussap_2021} based on the Laguerre expansion, where the coefficients $\{a^\phi_k\}_{k\in\mathbb{N}}$ for the expansion $\phi_m$ are not obtained via the aforementioned recurrence formula, but directly via the Plancherel theorem.
This approach eases the computation required for estimation, whereas it then seems difficult to explore the limiting distribution of the estimator, that is regarded as a benefit from employing the Laguerre series expansion in this context.


%% file: sections_ver2x/edpf_concludingremarks.tex

\section{Concluding remarks}\label{section concluding remarks}

With a view towards further application of the Gerber-Shiu function in practice, the present survey aims to enhance research activities on not fully explored implementation aspects by casting fresh light on its numerical and statistical aspects for further developments.
To this end, we have provided a systematic summary on the formulation of the Gerber-Shiu function and its variants, the risk surplus models and additional structural features (Section \ref{section the gerber-shiu discounted penalty function}), as well as a comprehensive up-to-date survey on a wide variety of analytical, semi-analytical and asymptotic methods (Section \ref{section analytical and semi-analytical approaches}). 
Built on those essential background materials, we have grouped implementation aspects of the Gerber-Shiu function into the aforementioned two categories, namely, numerical methods (Section \ref{section numerical methods}) and statistical inference (Section \ref{section inference}), along with further classifications and representative concepts and formulas without going into too much technical detail to avoid digressing from the main objective as a comprehensive survey.

In contrast to its richness and great versatility, the Gerber-Shiu function does not seem sufficiently prevalent in practice.
As the present survey has revealed along the way, a direct and critical reason lies in a variety of difficulties in numerical approximation and statistical inference, particularly when its formulation and underlying risk models become more complex for capturing more features.
For instance, by making full use of its form of the expectation functional, Monte Carlo methods can be applied to every formulation, whereas the computing time required for decent accuracy on a practical standard can often be prohibitive, particularly in the rare event situation due to the need for generating many, possibly almost endlessly many, sample paths all the way up to their ruins.
To the best of our knowledge, even typical improvements, such as variance reduction techniques and sequential stopping rules, have not been explored fully in a specialized manner to the Gerber-Shiu function as of yet, and thus can be promising future research topics to attend to.  
As such, there should still be large room for development and improvement on numerical and statistical methods, in fact, out of necessity if more practical use of the Gerber-Shiu function is truly strived for.
We hope that the present comprehensive survey opens the door wide in such research directions as a useful manual of selecting and even improving the appropriate numerical and statistical approach for relevant risk models with suitable structural features.

Another, and equally crucial, cause for the Gerber-Shiu function not getting a deserved amount of practical attention is a lack of direct research interest in its practical use per se.
Despite its generality has long been well and widely understood, it is not as trivial for all interested parties to further translate real-world problems in the risk management into the Gerber-Shiu function.
There does not seem to be an active line of research in the literature so far addressing how to set the penalty function to real-world problems in the insurance industry.
Far beyond the aforementioned dividend and capital injection problems, the all-round form of the Gerber-Shiu function should lend itself to address a much wider variety of important practical problems, for instance, in the solvency risk assessment (such as debt reserves and margin ratios), the credit risk management, and applications as risk measures (far broader than the traditional framework of the ruin probability), just to name a few.
Such practical studies would almost certainly drive up the entire research activities around the Gerber-Shiu function in various directions.

Last but not least, the coding of numerical and statistical methods is generally not a trivial task, especially for realistic large-scale problems in the industry.
For more practical use and deserved dissemination of the Gerber-Shiu function, it is undoubtedly effective and even necessary to somehow promote the sharing of packaged codes (for instance, in Python or R) in appropriate source code repositories, such as GitHub, for free public use.